\documentclass[a4paper,11pt]{article}
\pdfoutput=1 

\usepackage{jheppub} 
\usepackage{color}
\usepackage{graphicx}
\usepackage{amsmath,latexsym}
\usepackage{amssymb}
\usepackage{xspace}
\usepackage[small]{subfigure}
\usepackage{xcolor}
\usepackage{appendix}
\graphicspath{{figures/}}

\title{Phase space path integral approach to the kinetics of black hole phase transition}

\author[a]{Ran Li,}
\author[b]{Conghua Liu,}
\author[c,*]{Jin Wang \note[*]{Corresponding authors.}}
\affiliation[a]{Department of Physics, Qufu Normal University, Qufu, Shandong 273165, China}
\affiliation[b]{State Key Laboratory of Electroanalytical Chemistry, Changchun Institute of Applied Chemistry, Chinese Academy of Sciences, Changchun 130022, China}
\affiliation[c]{Department of Chemistry, and Department of Physics and Astronomy, The State University of New York at Stony Brook, Stony Brook, NY 11794, USA}

\emailAdd{liran@qfnu.edu.cn}
\emailAdd{jin.wang.1@stonybrook.edu}

\abstract{We employ the approach of path integral in the ``phase space" to study the kinetics of state switching associated with black hole phase transitions. Under the assumption that the state switching process of the black hole is described by the stochastic Langevin equation based on the free energy landscape, we derived the Martin–Siggia–Rose–Janssen–de Dominicis (MSRJD) functional and obtained the path integral expression of the transition probability. The MSRJD functional inherently represents the path integral in the ``phase space", allowing us to extract the effective Hamiltonian for the dynamics of state switching process. By solving the Hamiltonian equations of motion, we obtain the kinetic path in the ``phase space'' using an example of the RNAdS black hole. Furthermore, the dominant kinetic path within the configuration space is calculated. We also discuss the kinetic rate by using the functional formalism. Finally, we examine two further examples: Hawking-Page phase transition and Gauss-Bonnet black hole phase transition at the triple point. Our analysis demonstrates that, concerning the Hawking-Page phase transition, while a dominant kinetic path in the ``phase space" from the large SAdS black hole to the thermal AdS space is present, there is no kinetic path for the inverse process. For the Gauss-Bonnet black hole phase transition at the triple point, the state switching processes between the small, the intermediate and the large Gauss-Bonnet black holes constitute a ``chemical reaction cycle". }
\begin{document}

\maketitle
\section{Introduction}

Since the discovery of black hole thermodynamics \cite{Hawking:1975vcx}, it has become widely accepted that black holes serve as an ideal model for exploring the profound connections between general relativity, quantum theory, and statistical physics. Concepts and methods from the realms of quantum theory and statistical physics have found extensive application in our quest to unravel the nature of gravity and black holes. Partly driven by the AdS/CFT correspondence \cite{Maldacena:1997re}, the investigation of black hole thermodynamics in anti-de Sitter (AdS) space, especially the phase transition of the AdS black holes, has attracted significant attention. Hawking and Page's seminal work revealed a first-order phase transition between the Schwarzschild AdS black hole and thermal AdS space \cite{Hawking:1982dh}, subsequently interpreted as the confinement/deconfinement transition of the dual quark-gluon plasma \cite{Witten:1998qj}. More recently, the study of black hole phase transitions in AdS space has been further propelled by the idea that the cosmological constant can be regarded as the thermodynamic pressure \cite{Kastor:2009wy,Dolan:2011xt}. Within this expanded phase space framework, AdS black holes have been shown to exhibit a wide array of fascinating phenomena, including behaviors reminiscent of Van der Waals fluids \cite{Kubiznak:2012wp}, re-entrant transitions \cite{Altamirano:2013ane}, the existence of a triple point \cite{Altamirano:2013uqa}, superfluid characteristics \cite{Hennigar:2016xwd}, the proposal of a molecular interpretation for the fundamental constituent degrees of freedom \cite{Wei:2015iwa}, and insights into the microstructure of black holes \cite{Wei:2019uqg}. This specialized field of study is commonly referred to as "black hole chemistry."

More recently, considerable efforts have been dedicated to investigating the kinetics of state switching associated with the black hole phase transitions by using the method of stochastic dynamics \cite{Li:2020khm,Li:2020nsy}. The starting point of this approach is to consider that AdS black holes interact with the thermal environment (thermal bath) located at the boundary of AdS space. This interaction leads to an exchange of energy with the environment, inducing stochastic changes of the black hole event horizon in the bulk. The black holes, formed through this process, are referred to as "fluctuating black holes" and are treated as the intermediate states in the state switching process. Originally, by introducing the event horizon radius as the order parameter to characterize the microscopic degrees of freedom of the fluctuating black hole in a coarse-grained way, the generalized free energy function of the fluctuating black hole is defined by using the thermodynamic relations. However, further studies show that the generalized free energy can be derived from the Euclidean gravitational action by employing the path integral method \cite{Li:2022oup}. With the generalized free energy in hand, one can intuitively exhibit the thermodynamic stability of the black holes by examining the free energy landscape, which is the pictorial representation of the generalized free energy as the function of the order parameter. The thermodynamics of black hole phase transition can be fully captured by the free energy landscape. It is further conjectured that the process of state switching associated with the black hole phase transition on the free energy landscape bears an analogy to the dynamical process of the Brownian particle driven by the potential. In this way, the stochastic Langevin equation has been proposed to describe the dynamics of state switching process associated with the black hole phase transition, where the deterministic part of the driven force is provided by the gradient force form the generalized free energy function while the stochastic force originates from the thermal fluctuations of the environment \cite{Li:2021vdp}.

In this context, the transition rate and the mean first passage time, as the characteristic quantities describing the kinetics of state switching process associated with the black hole phase transition, have been subjected to both analytical and numerical investigations \cite{Li:2021vdp}. Depending upon the damping coefficient, the kinetic turnover was discovered for the RNAdS black hole phase transition. It is believed that the occurrence of a kinetic turnover represents a universal feature in this type of phase transition. Furthermore, the relative fluctuation of the mean first passage time is shown to be a monotonic function of the damping coefficient and has a sudden change at the turnover point, which is proposed to characterize the microstructure of the black holes.

Notably, the Onsager-Machlup (OM) path integral formalism was employed to determine the most probable (dominant) kinetic path of black hole state switching process and to calculate the transition probability of the black holes \cite{Liu:2021lmr}. However, the OM functional can only give the path integral in the configuration space (the space composed by the order parameters). It is well known that in the non-equilibrium statistical physics, there is an alternative approach to study the stochastic Langevin dynamics, the Martin–Siggia–Rose–Janssen–de Dominicis (MSRJD) functional integral \cite{Martin:1973pra,Janssen:1976zpb,Dominicis:1978prb}. In fact, the relationship between the MSRJD path integral and the OM path integral is analogous to the one between the Hamiltonian formulation in the phase space and the Lagrangian formulation in the configuration space for the path integral in quantum mechanics \cite{Justin,Wio}. In order to get a deep understanding on the path integral approach, we exploit the MSRJD functional integral to study the kinetics of state switching process associated with black hole phase transition.

Starting with the Langevin equation that determined the stochastic evolution of the order parameter for the RNAdS black hole, we will give the derivation of the MSRJD functional, from which we can extract the effective Hamiltonian. We then study the Hamilton flow lines by solving the equation of motion in the phase space. According the physical picture of the black hole phase transition, we can determine the dominant kinetic path in the phase space. Finally, we compute the transition rate by evaluating the path integral through the saddle approximation method and discuss its relationship with the previous results \cite{Li:2021vdp,Liu:2021lmr}. This procedure will be further applied to study the kinetics of state switching associated with Hawking-Page phase transition and the Gauss-Bonnet AdS black hole phase transition.

This paper is organized as follows. In Sec.\ref{Sec_II}, we review the basic proposal of stochastic method to study the kinetics of black hole state switching process by taking the RNAdS black hole as an example. We also provide an alternative derivation of the generalized free energy by using the York's formalism. In Sec.\ref{Sec_III}, we present the derivation of the MSRJD functional integral for our problem. In Sec.\ref{Sec_IV}, we discussed the Hamiltonian flows and the transition rates. In Sec.\ref{Sec_V}, two further examples, the Hawking-Page phase transition and the Gauss-Bonnet AdS black hole phase transition are considered. The discussion and conclusion are presented in the last section.

\section{Stochastic dynamics of state switching associated with black hole phase transition on the free energy landscape}
\label{Sec_II}

\subsection{Free energy landscape of black hole}

The physical picture of state switching associated with black hole phase transition can be described as follows. Imagine that there is a black hole in spacetime and place a thermal bath somewhere. For the AdS black holes that we are mainly interested in this paper, the thermal bath can be located at the spacial infinity. Recall that a photon when radiated to spacial infinity can return in a finite proper time. Although the thermal bath is placed at the spacial infinity, the black hole can reach the equilibrium state with the thermal bath due to the interaction between them. When studying the black hole phase transition from the thermodynamic perspective, we realized that there may exist more than one branch of black hole solution. Different types of black holes can be distinguished by their radii of the event horizon \cite{Kubiznak:2012wp,Wei:2015iwa}. Due to the thermal fluctuations, the radius of the black hole event horizon can increase or decrease by absorbing or releasing the energy in a completely stochastic manner. The black hole phase transition process that a black hole starts from an equilibrium state and finally reaches another equilibrium state can be described well by the process of changing the black hole radius. Therefore, the state switching process associated with phase transition can be treated as the stochastic process caused by the thermal fluctuations from the bath. The radius of the black hole event horizon can be selected to be the order parameter to characterize the process of the state switching.

The order parameter is a fundamental concept in the study of phase transitions, critical phenomena, and symmetry breaking in physics. It serves as a key indicator of the system's state and provides valuable insights into the organization and behavior of a physical system. The order parameter is often related to a thermodynamic potential, such as the free energy or Landau-Ginzburg-Wilson free energy, which can be used to describe the phase transition \cite{Landau}. The forms of these potentials also depend on the particular order parameter associated with the phase transition.

For the black hole phase transition, the thermodynamic potential is just the free energy, which can be calculated from the gravitational action by using the Euclidean path integral method \cite{Gibbons:1976ue}. In this aspect, we have shown that the generalized free energy for the fluctuating black hole in Einstein gravity \cite{Li:2022oup} and Gauss-Bonnet gravity \cite{Li:2023men} can be readily derived by such a method. By plotting the generalized free energy as the function of the black hole radius, one can get the so-called free energy landscape, from which one can easily read off the stability of the black hole.

As an important example, let us review the van der Waals type phase transition of the RNAdS black hole in the extended phase space from the perspective of the free energy landscape \cite{Li:2020nsy}. RNAdS black hole is a solution to the Einstein-Maxwell gravitational theory. The metric is given by 
\begin{eqnarray}\label{RNAdS_metric}
ds^2=-f(r)dt^2+\frac{1}{f(r)}dr^2+r^2d\Omega_2^2\;,
\end{eqnarray}
with 
\begin{eqnarray}
    f(r)=1-\frac{2M}{r}+\frac{Q^2}{r^2}+\frac{r^2}{L^2}\;,
\end{eqnarray}
where $M$ is the mass, $Q$ is the electric charge, and $L$ is the AdS curvature radius. The electromagnetic gauge potential is given by 
\begin{eqnarray}
    A=-\frac{Q}{r} dt\;.
\end{eqnarray}

The event horizon is the largest positive root of the equation $f(r)=0$. According to this equation, the mass of the fluctuating black hole that is generated during the state switching process can be expressed as the function of the order parameter $r_+$ 
\begin{eqnarray}\label{mass}
    M=\frac{r_+}{2}\left(1+\frac{r_+^2}{L^2}+\frac{Q^2}{r_+^2}\right)\;.
\end{eqnarray}
The Bekenstein-Hawking entropy and the Hawking temperature are given by using the conventional way as 
\begin{eqnarray}
    S&=&\pi r_+^2\;,\label{entropy}\\
    T_H&=&\frac{1}{4\pi r_+}\left(1+\frac{3r_+^2}{L^2}- \frac{Q^2}{r_+^2}\right)\;.
\end{eqnarray}
For the fluctuating black hole, these quantities are also the functions of the order parameter $r_+$.

We now employ the Gibbons-Hawking path integral approach to study and interpret the thermodynamics of the fluctuating black hole. In the Gibbons-Hawking approach to black hole thermodynamics, the partition function of the canonical ensemble can be written in the form of the gravitational path integral \cite{Gibbons:1976ue}
\begin{eqnarray}\label{partition_func}
Z_{grav}(\beta)=\int D[g] e^{-I_E[g]}\simeq  e^{-I_E[g]} \;, 
\end{eqnarray}
where $I_E[g]$ is the Euclidean gravitational action and $\beta$ is the integral period of the Euclidean time. The Euclidean gravitational action for the Einstein-Maxwell theory is given by \cite{Emparan:1999pm,Caldarelli:1999xj}
\begin{eqnarray}\label{E_H_action}
I_E=I_{bulk}+I_{surf}+I_{ct}\;,
\end{eqnarray}
with
\begin{eqnarray}
    I_{bulk}&=&-\frac{1}{16\pi} \int_{\mathcal{M}}d^4x\sqrt{g} \left( R +\frac{6}{L^2}-F_{\mu\nu}F^{\mu\nu}\right)\;,\\
    I_{surf}&=&-\frac{1}{8\pi} \int_{\partial\mathcal{M}} d^3 x\sqrt{h} \left(K+2n_{\mu} F^{\mu\nu} A_{\nu}\right)\;,\\
    I_{ct}&=&\frac{1}{8\pi} \int_{\partial\mathcal{M}} d^3 x
    \sqrt{h}\left[\frac{2}{L}+\frac{L}{2}\mathcal{R} -\frac{L^3}{2} \left( \mathcal{R}_{ab}\mathcal{R}^{ab}-\frac{3}{8}\mathcal{R}^2\right) \right]\;.
\end{eqnarray}
Here, $h_{\mu\nu}=g_{\mu\nu}-n_{\mu}n_{\nu}$ is the induced metric on the boundary $\partial\mathcal{M}$ where $n$ is an outward pointing unit normal vector to $\partial\mathcal{M}$. In the Gibbons-Hawking boundary term $I_{surf}$, the trace of the extrinsic curvature is defined by $K=\nabla^{\mu}n_{\mu}$. In the counterterm action $I_{ct}$, $\mathcal{R}$ and $\mathcal{R}_{ab}$ are the Ricci scalar and Ricci tensor for the boundary metric, respectively. The Gibbons-Hawking surface term is to preserve a well posed variation problem while the surface term for the electromagnetic field is to preserve the fixed charge boundary condition at the infinity. The counterterm action is sufficient to cancel divergences.  

However, our calculation of the gravitational action is slightly different from the original Gibbons-Hawking approach \cite{Gibbons:1976ue}, where the time period $\beta$ is selected to preserve the regularity of the Euclidean geometry. Here, the period $\beta$ is an arbitrary parameter \cite{Susskind:1993ws,Carlip:1993sa,Fursaev:1994te,Susskind:1994sm}. Its physical meaning is the inverse temperature of the thermal bath. Then, this choice will introduce a conical singularity at the horizon. It is argued that the Euclidean geometry with the conical singularity is also the solution to the classical equations of motion \cite{Susskind:1994sm}. When evaluating the action functional, the horizon area or the horizon radius $r_+$ is kept fixed. This condition introduces a Lagrange multiplier, which results in the energy density along the $r=r_+$ surface in the phase space. The solution to the classical equations of motion will have a conical singularity at the horizon. Therefore, the Euclidean geometry with arbitrary $\beta$ is also a stationary point of the Euclidean functional integral as long as the constraint of the constant horizon radius is satisfied.

By evaluating the gravitational action and taking the conical singularity's contribution into account, one can get the Euclidean action for the fluctuating RNAdS black hole as 
\begin{eqnarray}\label{action}
I_{E}= \frac{\beta r_+}{2}\left(1+\frac{r_+^2}{L^2}+\frac{Q^2}{r_+^2}\right)-\pi  r_+^2\;.
\end{eqnarray}
The detailed calculation is presented in Appendix \ref{App_A}. According to the relationship between the thermodynamic potential and partition function of the canonical ensemble, one can obtain the generalized free energy function of the fluctuating RNAdS black hole, which is given by \begin{eqnarray}\label{GibbsEq}
F=-T\ln Z_{grav}=\frac{I_E}{\beta}=\frac{r_+}{2}\left(1+\frac{8}{3}\pi P r_+^2+\frac{Q^2}{r_+^2} \right)-\pi T r_+^2\;,
\end{eqnarray}
where the effective thermodynamic pressure $P=\frac{3}{8\pi}\frac{1}{L^2}$ is introduced, with $L$ being the AdS curvature radius \cite{Kastor:2009wy,Dolan:2011xt}. From the generalized free energy, one can also obtain the energy and the entropy of the fluctuating RNAdS black hole by using the conventional formulas in statistical physics \cite{Li:2022oup}.

By treating the black hole radius $r_+$ as the independent argument, one can formulate the generalized free energy landscape of the RNAdS black hole at the specific ensemble temperature $T$. The free energy landscape of the RNAdS black holes has the shape of double well in certain temperature regime \cite{Li:2020nsy}. The local stable state condition $\partial F/\partial r_+=0$ leads to three extreme points, i.e. three branches of the RNAdS black holes. They correspond to the small, the intermediate, and the large RNAdS black holes. In \cite{Li:2020nsy}, the stability of the three branches of RNAdS black holes was discussed based on the free energy landscape in detail. In addition, the condition for the local stable state will lead to the condition that the ensemble temperature is equal to the Hawking temperature of the local stable black hole 
\begin{eqnarray}\label{RNAdS_Temp}
T=T_H=\frac{1}{4\pi r_+}\left(1+8\pi P r_+^2- \frac{Q^2}{r_+^2}\right)\;.
\end{eqnarray}
When this condition is satisfied (we also refer this condition as the on-shell condition), the Euclidean geometry of the local stable RNAdS black hole is free of the conical singularity. In addition, the equation (\ref{RNAdS_Temp}) should be treated as the equation of state for the local stable RNAdS black hole \cite{Kubiznak:2012wp}
\begin{eqnarray}
P=\frac{T}{2r_+}-\frac{1}{8\pi r_+^2}+\frac{Q^2}{8\pi r_+^4}\;.
\end{eqnarray}
At the arbitrary ensemble temperature, the generalized free energy (\ref{GibbsEq}) is off-shell. If substituting the equation (\ref{RNAdS_Temp}) into the expression Eq.(\ref{GibbsEq}) of the generalized free energy, one can obtain the on-shell value of the free energy for the local stable RNAdS black hole \cite{Chamblin:1999hg,Chamblin:1999tk,Caldarelli:1999xj}.

We have to point out that the black hole radius $r_+$ as the order parameter of the small/large RNAdS black hole phase transition has a constraint from the non-negativity of the Hawking temperature $T_H$ for the local stable black hole. The constraint on the order parameter is then given by 
\begin{eqnarray}
r_+\geq \frac{L}{\sqrt{6}}\left(\sqrt{1+\frac{12Q^2}{L^2}}-1\right)^{1/2}\;. 
\end{eqnarray}
This constraint implies that the black hole cannot have arbitrary small event horizon due to the presence of electric charge \cite{Liu:2021lmr} or rotation \cite{Yang:2021ljn}.

\subsection{Generalized free energy from York's approach} 

Before discussing the stochastic dynamics of the black hole state switching, we provide an alternative derivation of the generalized free energy by using York's formalism \cite{York:1986it,Whiting:1988qr,Braden:1990hw}. In this formalism, the black hole is placed in a cavity with the fixed boundary conditions that can constitute a thermodynamic ensemble. However, in evaluating the gravitational action, only the constraint equations are satisfied, but not the full dynamical equations of motion. This results in the off-shell configurations that can be reached under the fluctuations.

In fact, the York's formalism has been applied to study the thermodynamics of the RNAdS black hole \cite{Brown:1994gs,Peca:1998cs,Mitra:1999ge}. The black hole is enclosed by a cavity with finite radius $r_B$. In this approach, one can calculate the action of the RNAdS black hole enclosed by cavity with the temperature $\beta$ and the radius $r_B$ as fixed parameters, and finally take the $r_B\rightarrow\infty$ limit to derive the generalized free energy for the case that the thermal bath is located at spatial infinity.

It is shown in \cite{Mitra:1999ge} that the resulting gravitational action for the RNAdS black hole in the canonical ensemble is given by
\begin{eqnarray}
    I=-\tilde{\beta} r_B \sqrt{1-\frac{r_+}{r_B}-\frac{r_+^3}{L^2 r_B} -\frac{Q^2}{r_+ r_B}+\frac{Q^2}{r_B^2}+\frac{r_B^2}{L^2}}+I_0-\pi r_+^2\;,
\end{eqnarray}
where $\tilde{\beta}$ is the inverse temperature of the cavity, $r_+$ is the horizon radius, $r_B$ is the radius of the cavity, and $I_0$ is an arbitrary term that can be used to define the zero of the energy. By choosing the thermal energy of the AdS space as the zero of energy, the authors in Ref.\cite{Peca:1998cs} argued that 
\begin{eqnarray}
    I_0=\tilde{\beta} r_B \sqrt{1+\frac{r_B^2}{L^2}}\;.
\end{eqnarray}
This can also be achieved by adding the counterterm term as shown in Appendix \ref{App_A}
\begin{eqnarray}
   I_{ct}=\frac{\tilde{\beta}}{2} r_B^2  \left[\frac{2}{L}+\frac{L}{r_B^2}-\frac{L^3}{4r_B^4} \right]\;,
\end{eqnarray}
where we have used the definition of inverse temperature at the cavity $\tilde{\beta}=2\pi b(r_B)$. It is easy to see that $I_0$ and $I_{ct}$ have the same asymptotic behavior.

Although the temperature of the cavity is treated as a fixed parameter, to keep the local thermodynamic equilibrium, or to eliminate the conical singularity of the Euclidean geometry, the temperature at the cavity is chosen to be the Hawking temperature blue-shifted to the cavity’s position $r_B$ \cite{York:1986it,Whiting:1988qr,Braden:1990hw}. For the on-shell RNAdS black hole, the relation is naively given by 
\begin{eqnarray}
    \tilde{\beta}=\beta_{H} \sqrt{1-\frac{2M}{r_B}+\frac{Q^2}{r_B^2}+\frac{r_B^2}{L^2}}\;.
\end{eqnarray}

For our concern of the RNAdS black hole with the thermal bath locating at spatial infinity, one should put the cavity to infinity by taking the limit $r_B\rightarrow +\infty$. It is clear that taking the limit gives a divergent inverse temperature $\tilde{\beta}$, which is unphysical. This can be fixed by noting that the black hole thermodynamics in AdS is infact defined on the conformal boundary. We should rescale the inverse temperature $\tilde{\beta}$ by a factor $\frac{L}{r_B}$, i.e. we can define the rescaled inverse temperature as
\begin{eqnarray}
    \beta=\frac{L}{r_B}\tilde{\beta}\;.
\end{eqnarray}
Taking the limit $r_B\rightarrow +\infty$, one can finally get the generalized free energy as 
\begin{eqnarray}
   I= \frac{\beta r_+}{2}\left(1+\frac{r_+^2}{L^2}+\frac{Q^2}{r_+^2}\right)-\pi  r_+^2\;,
\end{eqnarray}
which is consistent with the result given in Eq.\eqref{action}.

In deriving this result, only the Gauss-law constraints for the electromagnetic field and the Hamiltonian constraint for the gravitational field are used \cite{Peca:1998cs,Mitra:1999ge}. However, the full Einstein equations are not required to satisfy. This inconsistency can be regarded as a kind of quantum fluctuations near the stationary black hole solution. In this sense, the generalized free energy is considered to describe the fluctuating off-shell black holes as well.

\subsection{Stochastic Langevin equation for the black hole state switching}

\begin{figure}
    \centering
    \includegraphics[width=0.45\textwidth]{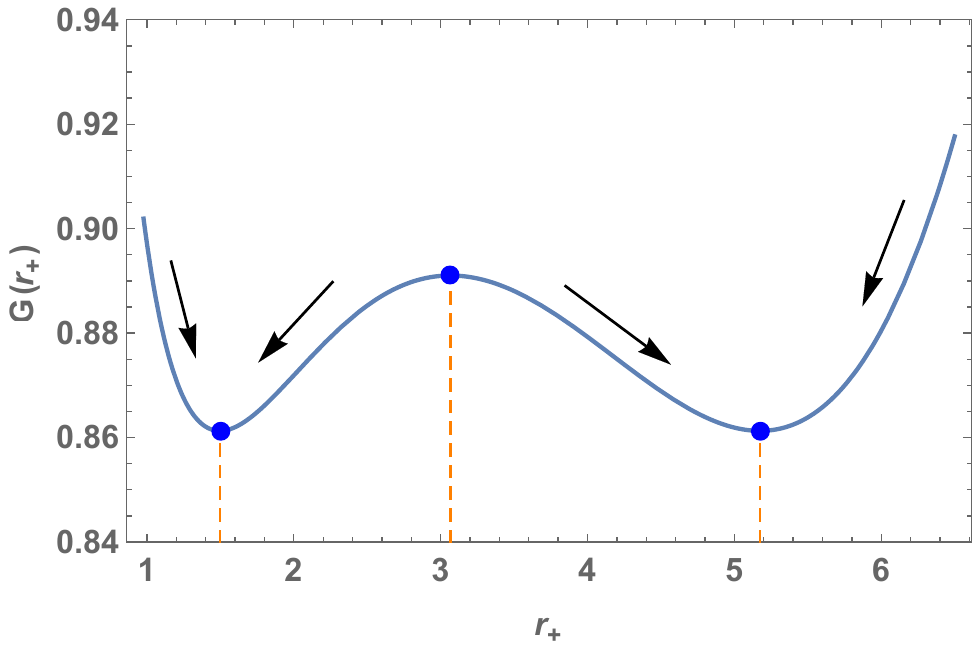}\;\;\;
    \includegraphics[width=0.45\textwidth]{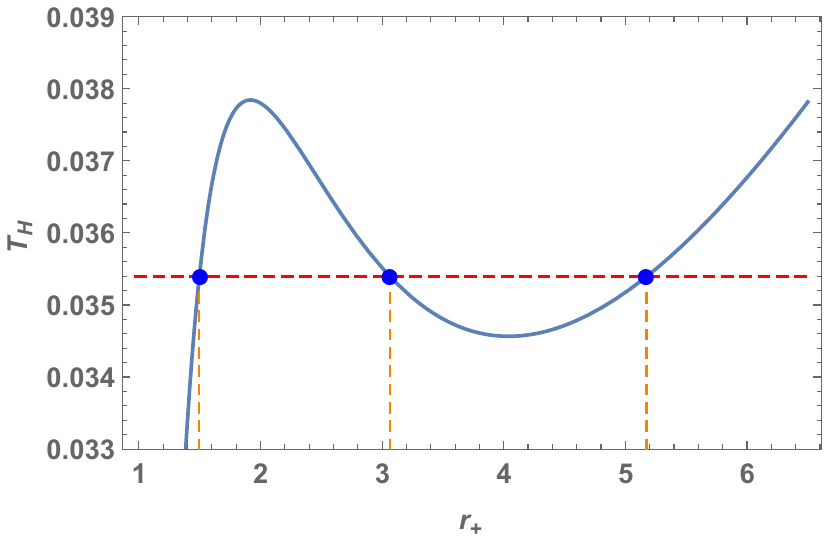}
    \caption{Left: Generalized free energy function at the phase transition point. Right: Hawking temperature as the function of black hole radius. In the two plots, the three blue points represent the three branches of RNAdS black hole solutions and the three vertical lines denote their radii $r_s,r_m$ and $r_l$. In the right plot, the horizontal red dashed line represents the temperature of the phase transition point. }
    \label{RNAdS_landscape}
\end{figure}

To give a physical picture of the above statements, let us discuss the free energy landscape for the RNAdS black hole phase transition. In Figure \ref{RNAdS_landscape}, we plot the free energy landscape and the Hawking temperature of the fluctuating RNAdS black holes. It should be noted that the order parameter, i.e. the horizon radius $r_+$, is taken to be the independent argument. On the left plot, the ensemble temperature is selected to be the phase transition temperature, at which the small black hole and the large black hole coexist. On the landscape, the three blue dots represent the three branches of the RNAdS black holes. For convenience, we denote the radius of the small/medium/large RNAdS black hole as $r_s/r_m/r_l$. They are the extremal points on the landscape. It can be seen that the small and the large RNAdS black holes have the same free energy. On the right plot, the Hawking temperature $T_H$ of the fluctuating black hole is plotted as the function of the order parameter $r_+$. The horizontal line represents the ensemble temperature $T$, which has been selected to be the phase transition temperature.    

We now consider the deterministic relaxation process of the fluctuating RNAdS black holes without the fluctuating force. In the left panel of Figure \ref{RNAdS_landscape}, we have explicitly indicated the the directions of these processes. As an example, we consider a fluctuating RNAdS black hole with the horizon radius $r_m<r_+<r_l$. From the right plot, we can see the Hawking temperature $T_H$ of this fluctuating black hole is smaller than the ensemble temperature. The deterministic relaxation process of this fluctuating black hole is just the absorbing energy process from the thermal bath, which leads to the increase in the energy itself. In the range of $r_+>r_0$, the mass of the RNAdS black hole is the monotonic increasing function of the radius. Therefore, increasing energy will in turn give rise to the increase of the black hole order parameter $r_+$. As a result, the fluctuating RNAdS black hole with radius $r_m<r_+<r_l$ will relax to the state of the large black hole with radius $r_+=r_l$ finally as indicated explicitly in the left plot. From the above analysis, we can conclude that the free energy landscape gives a physical picture of the deterministic relaxation process of the fluctuating black holes without the fluctuating forces.

However, this is not adequate to explore the kinetics of the black hole state switching. If only considering the deterministic gradient force from the free energy landscape, the local stable black hole can never make a transition and switch to another local or global stable black hole. It is the fluctuating force (thermal noise) from the bath that makes the state switching process possible. Therefore, the dynamics of the black hole state switching should be described by the Langevin equation for the stochastic evolution or equivalently the Fokker-Planck equation for the probabilistic evolution. Therefore, the methods from the non-equilibrium statistical physics and the stochastic dynamics can be applied to study the kinetics of state switching associated with the black hole phase transition \cite{ZNSM,Kampen}.

Much inspiration can be gained from the so-called time-dependent Ginzburg-Landau model \cite{Hohenberg}, which is briefly reviewed in the Appendix B. In this model, the collective behavior of the system near the critical point is described by Langevin equation. With these observations in mind, we can write down the Langevin equation that governs the dynamics of state switching process associated with the black hole phase transition as follows \cite{ZNSM,Kampen,Li:2021vdp}
\begin{eqnarray}
    \ddot{\phi}+\zeta \dot{\phi} +G'(\phi) -\tilde{\eta}(t)=0\;,
\end{eqnarray}
where $\phi$ denotes the order parameter $r_+$ and a dot denotes a derivative respect to $t$. The effective friction coefficient $\zeta$ is introduced to describe the interaction between the black hole and its environment. The stochastic noise $\tilde{\eta}(t)$ from the environment satisfies the relation 
\begin{eqnarray}
    \langle \tilde{\eta}(t) \rangle=0\;,\;\;\; \langle \tilde{\eta}(t)\tilde{\eta}(t') \rangle=2 \zeta T \delta(t-t')\;.
\end{eqnarray}
In the overdamped regime, the Langevin equation can be simplified as 
\begin{eqnarray}\label{overdamped_Langevin_eq}
    \dot{\phi}= -\frac{1}{\zeta}G'(\phi) +\eta(t)\;,
\end{eqnarray}
where $\eta(t)=\frac{1}{\zeta}\tilde{\eta}(t)$ is introduced for simplicity. It satisfies the fluctuating-disspertion relation 
\begin{eqnarray}
    \langle \eta(t)\eta(t') \rangle=2D\delta(t-t')\;,
\end{eqnarray}
where $D=\frac{T}{\zeta}$ is the diffusion coefficient. This is just the overdamped Langevin equation that describes the stochastic motion of the Brownian particle.

In the free energy landscape formalism, the generalized free energy is analogous to the Landau free energy. By introducing the black hole radius as the order parameter, the microscopic degrees of freedom of the black hole can be described collectively. In this way, the collective motion of the microscopic degrees of freedom of the black hole is reflected by the change in the order parameter. It is wisely accepted that in the ordinary mean field theory, especially near the critical point, the collective behavior of the system can be well described by the Langevin equation \cite{Hohenberg}. Inspired by this observation, we have introduced the stochastic Langevin equation to characterize the state switching process associated with the black hole phase transition.

However, in writing down the stochastic Langevin equation, the friction coefficient $\zeta$ or equivalently the diffusion coefficient $D$ was introduced effectively. They are assumed to describe the complex interaction between the black hole and the thermal bath. Here, we want to draw the reader's attention to the discussion of observation signature of the quantum fluctuations of spacetime in the gravitational wave interferometer \cite{Verlinde:2019xfb,Verlinde:2019ade} (see also \cite{Parikh:2020kfh}). By using the holographic assumption, it is shown that, similar to the random walk model, the accumulated fluctuation over a path of length $L$ is of the variance of $\langle \Delta L^2 \rangle\sim l_p L$, where $l_p$ is the Planck length. We may also gain some insights from the stochastic inflation model \cite{Starobinsky,Miyachi:2023fss}. In this model, by treating the short wavelength quantum modes as the stochastic noise, the Langevin equations that describe the dynamics of long wavelength modes can be derived from the equation of motion for the scalar field. Although the thermal fluctuations considered in our model are distinct from the quantum fluctuations, we can still believe that there exists a mechanism of deriving the effective friction coefficient or the diffusion coefficient from the fundamental perspective as for the stochastic inflation model. This is out of the scope of the present work. We will leave it for future study.

\section{MSRJD path integral representation for the stochastic dynamics of the black hole state switching}
\label{Sec_III}

In this section, we discussed the path integral representation of the stochastic dynamics of the black hole state switching. We provide the derivation of the MSRJD functional from the overdamped Langevin equation.

Let us discretize the time evolution of the order parameter and the stochastic noise as 
\begin{eqnarray}
    \phi(t)\rightarrow \phi_i\;,\;\;\;\eta(t)\rightarrow \eta_i\;,
\end{eqnarray}
where $i\in Z$ denotes the time step of the discretization. Then the overdamped Langevin equation can be written as 
\begin{eqnarray}
    \psi_i:=\phi_i-\phi_{i-1}+\Delta t \left(\frac{1}{\zeta} G'(\phi_i)-\eta_i\right)=0\;,
\end{eqnarray}
where $\Delta t$ is the time interval. This scheme is called the Ito discretization \cite{Kampen}. 

Consider the generating functional, which is analogous to the partition function in statistical physics. Denote a solution to the Langevin equation with the noise $\eta$ as $\phi[\eta]$. It is useful to notice the identity is satisfied 
\begin{eqnarray}
    1=\int \mathcal{D}\phi \delta(\phi-\phi[\eta])=\int \mathcal{D}\phi \left|\frac{\delta \psi}{\delta \phi} \right|\delta(\psi) \;, 
\end{eqnarray}
where $\mathcal{D}\phi=\prod_{i} d\phi_i$ is the functional integral measure, $\delta(\phi-\phi[\eta])=\prod_i(\phi_i-\phi_i[\eta])$ and $\left|\frac{\delta \psi}{\delta \phi} \right|$ is the determinant of the Jacobian matrix $\left\{ \frac{\partial \psi_i}{\partial \phi_j} \right\}$. In the Ito discretization, the Jacobian matrix is a triangular matrix with the unit as the diagonal components. Therefore, $\left|\frac{\delta \psi}{\delta \phi} \right|=1$. Then we can get
\begin{eqnarray}
    1=\int \mathcal{D} \phi \prod_i \delta\left(\phi_i-\phi_{i-1}+\Delta t \left(\frac{1}{\zeta} G'(\phi_i)-\eta_i\right)\right)\;.
\end{eqnarray}
By using the integral representation of the delta function and taking the continuum limit, we can rewrite the above equation as 
\begin{eqnarray}
    1=\int \mathcal{D}\phi \mathcal{D}\tilde{\phi} \exp \left[ -i\int dt
    \tilde{\phi} \left(\dot{\phi}+\frac{1}{\zeta}G'(\phi)-\eta(t) \right)
    \right]\;.
\end{eqnarray}
Then the generating functional for the Langevin dynamics in the overdamped regime can be given by 
\begin{eqnarray}
    \mathcal{W}=\mathcal{N}  
    \int \mathcal{D}\phi \mathcal{D}\tilde{\phi} \mathcal{D}\eta(t) \exp \left[-\frac{1}{4D}\int dt \eta^2(t) \right]
    \exp \left[-i\int dt\tilde{\phi} \left(\dot{\phi}+\frac{1}{\zeta}G'(\phi)-\eta(t) \right)
    \right]\;,
\end{eqnarray}
where $\mathcal{N}$ is the normalization constant and the average on the Gaussian noise is taken into account. By performing the Gauss integral for the noise, one can get
\begin{eqnarray}
    \mathcal{W}=\mathcal{N}  
    \int \mathcal{D}\phi \mathcal{D}\tilde{\phi} \exp \left\{-\int dt\left[i\tilde{\phi} \left(\dot{\phi}+\frac{1}{\zeta}G'(\phi)\right)+D\tilde{\phi}^2\right]
    \right\}\;.
\end{eqnarray}

The transition probability $p(\phi,t;\phi_0,t_0)$ with the initial boundary condition $\phi(t_0)=\phi_0$ and final boundary $\phi(t)=\phi$ is then given by \cite{Wio}
\begin{eqnarray}
    p(\phi,t;\phi_0,t_0)=\int_{\phi_0}^{\phi} \mathcal{D}\phi \mathcal{D}\tilde{\phi} \exp \left\{-\int_{t_0}^{t} dt'\left[i\tilde{\phi(t')} \left(\dot{\phi}(t')+\frac{1}{\zeta}G'(\phi(t'))\right)+D\tilde{\phi}^2(t')\right]
    \right\}\;,
\end{eqnarray}
where the prefactor $\frac{\mathcal{N}}{\mathcal{W}}$ is absorbed into the integral measure.

In the present work, we are interested in the Hamiltonian formulation of the stochastic path integral. By introducing the conjugate momentum $\Pi=i\tilde{\phi}$, the probability can be rewritten as 
\begin{eqnarray}\label{kinetic_rate_PI}
    p(\phi,t;\phi_0,t_0)=\int_{\phi_0}^{\phi} \mathcal{D}\phi \mathcal{D}\tilde{\phi} \exp \left[-\int_{t_0}^{t} dt' \left(\Pi \dot{\phi}-\mathcal{H}\right)  \right]\;,
\end{eqnarray}
where the effective Hamiltonian $\mathcal{H}$ for the dynamics of state switching is defined as 
\begin{eqnarray}\label{Hamiltonian}
    \mathcal{H}=D\Pi^2-\frac{1}{\zeta} G'(\phi) \Pi\;. 
\end{eqnarray}

In summary, we have derived the MSRJD functional that can be used to explore the kinetics of state switching process associated with the black hole phase transition. The further discussions will be presented in the next two sections.

Now we study some implications of the MSRJD functional. In the phase space path integral representation of the transition probability, the integration over the momentum $\Pi$ extends to the imaginary axis \cite{Altland}. In addition, the integration over the momentum $\Pi$ is quadratic, which allows us to integrate over $\Pi$ to obtain the path integral in the Lagrangian form. From Eq.\eqref{kinetic_rate_PI}, one can derive the Lagrangian formula for the stochastic path integral by performing the Gaussian integral of $\Pi$ \cite{Wio}
\begin{eqnarray}
    p(\phi,t;\phi_0,t_0)=\int_{\phi_0}^{\phi} \mathcal{D}\phi \exp \left\{-\frac{1}{4D}\int_{t_0}^{t} dt'\left(\dot{\phi}(t')+\frac{1}{\zeta}G'(\phi(t'))\right)^2
    \right\}\;,
\end{eqnarray}
which is just the Onsager-Machlup path integral used in \cite{Liu:2021lmr}.

The corresponding Lagrangian is given by
\begin{eqnarray}
    \mathcal{L}=\frac{1}{4D}\left(\frac{d\phi}{dt}\right)^2+\frac{1}{2T} G'(\phi) \frac{d\phi}{dt}+\frac{1}{4T\zeta} \left(G'(\phi)\right)^2\;,
\end{eqnarray}
which is analogous to the Lagrangian of a one-dimensional particle with the mass $\tilde{m}=\frac{1}{2D}$ and the unit charge $\tilde{q}=1$ moving in an external electromagnetic field $\mathcal{A}(t,x)=\frac{1}{2T}G'(x)$ and an external potential $\mathcal{V}(t,x)=-\frac{1}{4T\zeta} \left(G'(x)\right)^2$
\begin{eqnarray}
    \mathcal{L}=\frac{\tilde{m}}{2}\dot{x}^2+\tilde{q}\mathcal{A}(t,x) \dot{x}-\mathcal{V}(t,x)\;.
\end{eqnarray}

At last, we show an intuitive derivation of the Fokker-Planck equation that describes the time evolution of the probability distribution $P(t, \phi)$. By replacing the momentum $\Pi$ with the operator $-\frac{\partial}{\partial x}$, one can write down the expression of the Hamiltonian given by Eq.\eqref{Hamiltonian} in the operator form 
\begin{eqnarray}
    \hat{H}=
    D\frac{\partial^2}{\partial \phi^2}+\frac{1}{\zeta} \frac{\partial}{\partial \phi} G'(\phi)=D\frac{\partial}{\partial \phi} \left(e^{-\beta G(\phi)}\frac{\partial}{\partial\phi}e^{\beta G(\phi)}\right)\;.
\end{eqnarray}
Then the Fokker-Planck equation can be written as the form of the ``Schr\"{o}dinger equation" \cite{Kim:2023gaa}
\begin{eqnarray}
   \frac{\partial P(t,\phi)}{\partial t}=D\frac{\partial}{\partial \phi} \left[e^{-\beta G(\phi)}\frac{\partial}{\partial\phi}\left(e^{\beta G(\phi)} P(t,\phi)\right)\right]\;,
\end{eqnarray}
which is just the Fokker-Planck equation that used in \cite{Li:2020khm} to study the kinetics of Hawking-Page phase transition. The above discussions imply that the dynamics of the state switching process associated with the black hole phase transition based on the free energy landscape is essentially unified in the framework of the phase space path integral formalism.

\section{Hamiltonian flow lines in phase space and kinetic rate of the black hole state switching}
\label{Sec_IV}

In this section, we will analyze the Hamiltonian flow lines in the phase space for the state switching process of the RNAdS black hole by solving the equations of motion numerically.   

\subsection{Hamiltonian flow lines}

The Hamilton's equations of motion can be given by
\begin{eqnarray}
    \dot{\phi}&=&2D\Pi-\frac{1}{\zeta} G'(\phi)\;,\\
    \dot{\Pi}&=&+\frac{1}{\zeta} G''(\phi) \Pi\;.
\end{eqnarray}
In the phase space, the fixed points are determined by the equations
\begin{eqnarray}
    \dot{\phi}= \dot{\Pi}=0\;,
\end{eqnarray}
which gives us 
\begin{eqnarray}
    \begin{cases}
			G'(\phi)=0\\
            \Pi=0
	\end{cases}
 \;\;\textrm{or} \;\;\;
    \begin{cases}
			G''(\phi)=0\\
            \Pi=\frac{1}{2T}G'(\phi) 
	\end{cases}
\end{eqnarray}
The first set of equations implies that the stationary points on the landscape, which are the on-shell black hole states satisfying the condition $G'(\phi)=0$, are also the fixed points in the phase space. Therefore, the solutions to the first set of fixed point's equations correspond to the three branches of the RNAdS black hole. In addition, for the first set of equations, the corresponding Hamiltonian is zero. For the second set, $G'(\phi)\neq 0$ and the corresponding Hamiltonian is negative. The equation $G'(\phi)\neq 0$ implies that the solutions to the second set of the fixed point's equations are not relevant to our discussion on the switching processes between the local or the global stable states on the free energy landscape.

\begin{figure}
    \centering
    \includegraphics[width=0.45\textwidth]{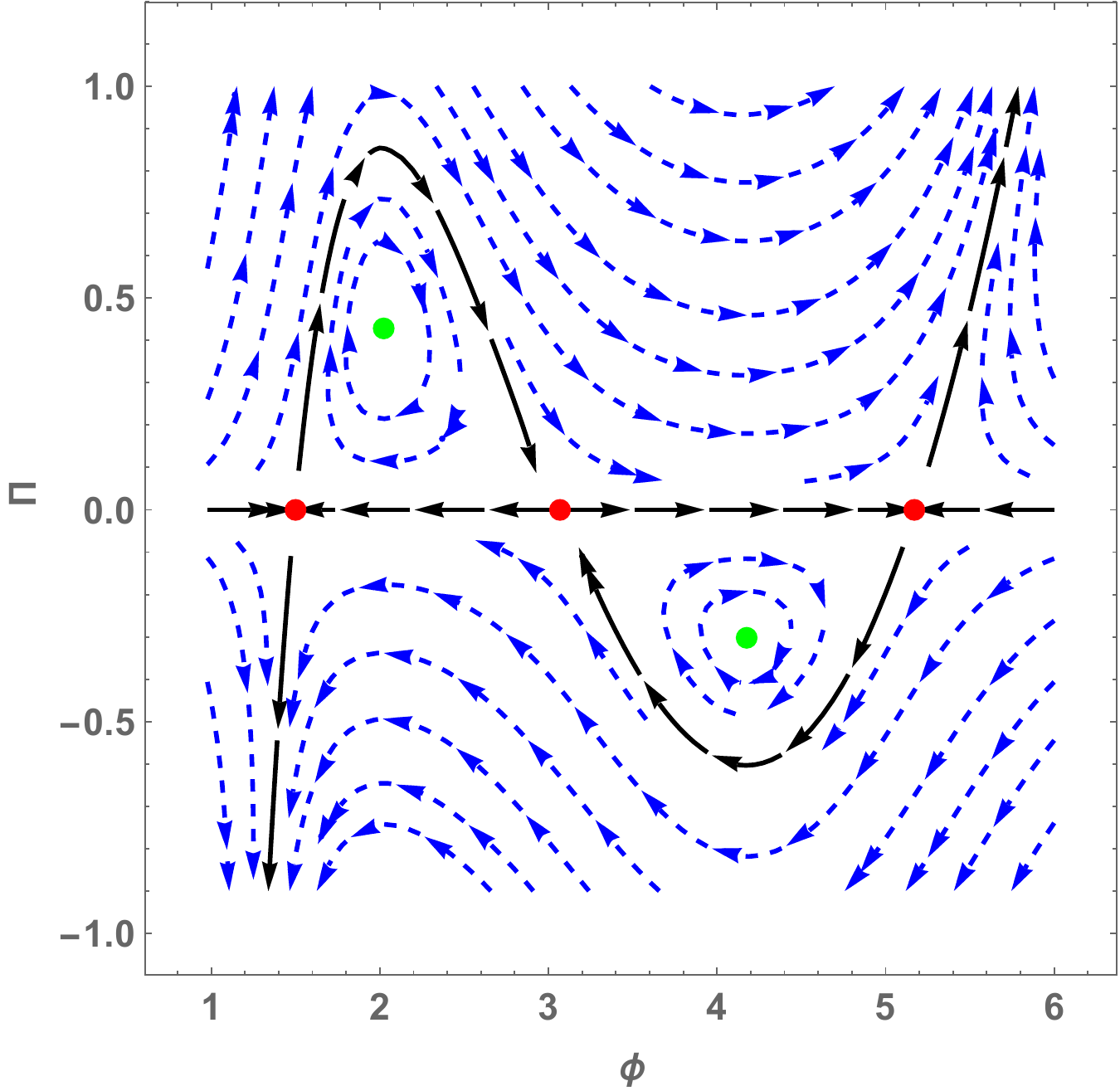}
       \caption{Hamiltonian flow lines or phase portrait for Hamilton’s equations of motion. The temperature is taken to be the phase transition temperature. There are five fixed points in the phase space. The red points are saddle points and the green points are centers. The saddle point has one unstable direction while the centers are not relevant to our analysis. The fixed points denoted by red color correspond the three branches of the RNAdS black hole solution. Each flow line can be specified by the value of Hamiltonian since it is conserved on the given flow line. The Hamiltonian vanishes along “zero-energy” lines (black lines). These lines connect
       the saddle points of Hamilton’s equations. }
    \label{Hamiltonian_flow_lines}
\end{figure}

Hamiltonian flow lines or phase portraits for Hamilton’s equations of motion are plotted in Figure \ref{Hamiltonian_flow_lines}. They reflect the stability of the fixed point intuitively. In this plot, the ensemble temperature is taken to be the phase transition temperature, where the two wells on the landscape have the same depth. The corresponding free energy landscape is shown in the left panel of Figure \ref{RNAdS_landscape}. In fact, the plot in Figure \ref{Hamiltonian_flow_lines} is typical as long as the free energy landscape has the shape of double well. Each flow line can be specified by a value of Hamiltonian since it is conserved on the given flow line. The “zero-energy” lines (black lines) where the Hamiltonian vanishes are given by 
\begin{eqnarray}
    \Pi=0\;,\;\;\;\textrm{or} \;\;\;\Pi=\frac{1}{T} G'(\phi)\;.
\end{eqnarray}
Along the $\Pi=0$ flow line, the Hamilton's equation becomes 
\begin{eqnarray}\label{Pieq0}
    \dot{\phi}=-\frac{1}{\zeta} G'(\phi)\;.
\end{eqnarray}
For the second case, $\Pi\neq 0$ and the Hamilton's equation becomes 
\begin{eqnarray}\label{Pineq0}
    \dot{\phi}=\frac{1}{\zeta} G'(\phi)\;.
\end{eqnarray}

On the $\Pi=0$ flow line, the equation for the order parameter $\phi$ describes the averaged equation of motion for the Langevin dynamics \eqref{overdamped_Langevin_eq} by noting that 
\begin{eqnarray}
    \langle \dot{\phi} \rangle =-\frac{1}{\zeta} \langle G'(\phi)\rangle\simeq -\frac{1}{\zeta} G'(\langle \phi \rangle)\;.
\end{eqnarray}
The topology of the phase space is determined by the “zero-energy” Hamiltonian flow lines. The two “zero-energy” flow lines intersect at the fixed points that represent the three branches of black hole solutions. On the $\Pi=0$ flow line, the state of the small RNAdS black hole and the state of the large RNAdS black hole are stable, while the state of the intermediate black hole on the top of free energy potential is unstable. On the flow line with $\Pi\neq 0$, the stability is reversed. Here, it should be distinguished that the stability of the black hole in the phase space reflects its dynamical nature while the stability of the black hole on the free energy landscape reflects the thermodynamic nature.

We denote the radii of the three branches of the RNAdS black hole as $r_s$, $r_m$ and $r_l$, respectively. The tunneling configuration can be easily obtained by analyzing the Hamiltonian flow lines. The tunneling configuration from the small black hole to the large black hole is starting from the point $(r_s,0)$, going to the point $(r_m,0)$ along the “zero-energy” flow line with $\Pi\neq 0$, and finally going to the point $(r_l,0)$ along the flow line with $\Pi=0$. It should be noted that, there are also an inverse tunneling configuration that represents the transition from the large black hole to the small black hole. This configuration in the phase space is starting from the point $(r_l,0)$, going to the point $(r_m,0)$ along the flow line with $\Pi\neq 0$, and finally arriving at the point $(r_s,0)$ along the flow line with $\Pi=0$. By numerically solving the Hamiltonian equations, one can obtain the corresponding tunneling configuration for the state switching process. This will be illustrated later.

\begin{figure}
    \centering
    \includegraphics[width=0.45\textwidth]{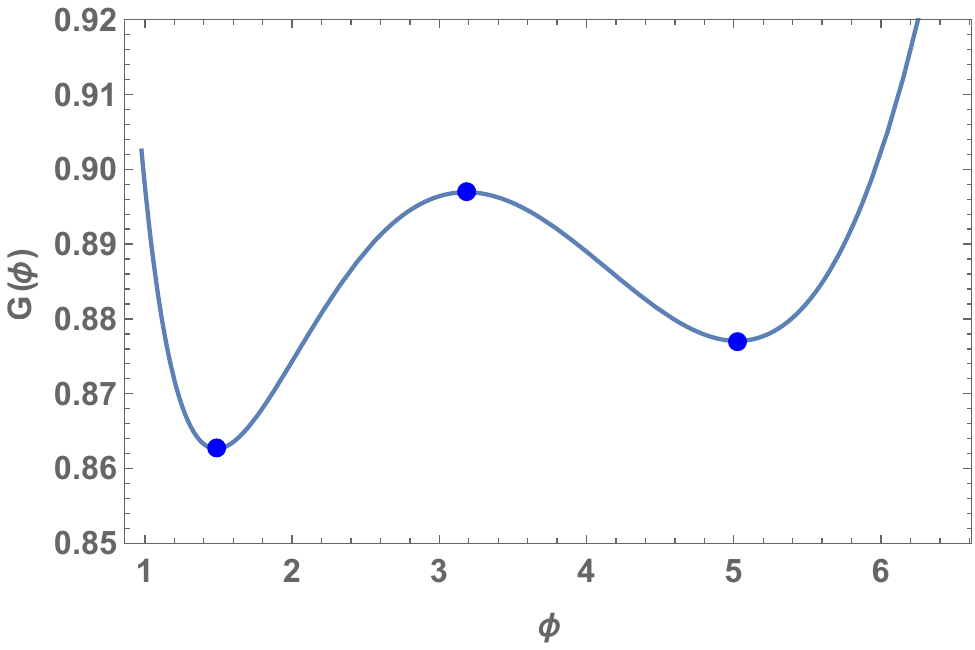}\;\;\;
    \includegraphics[width=0.45\textwidth]{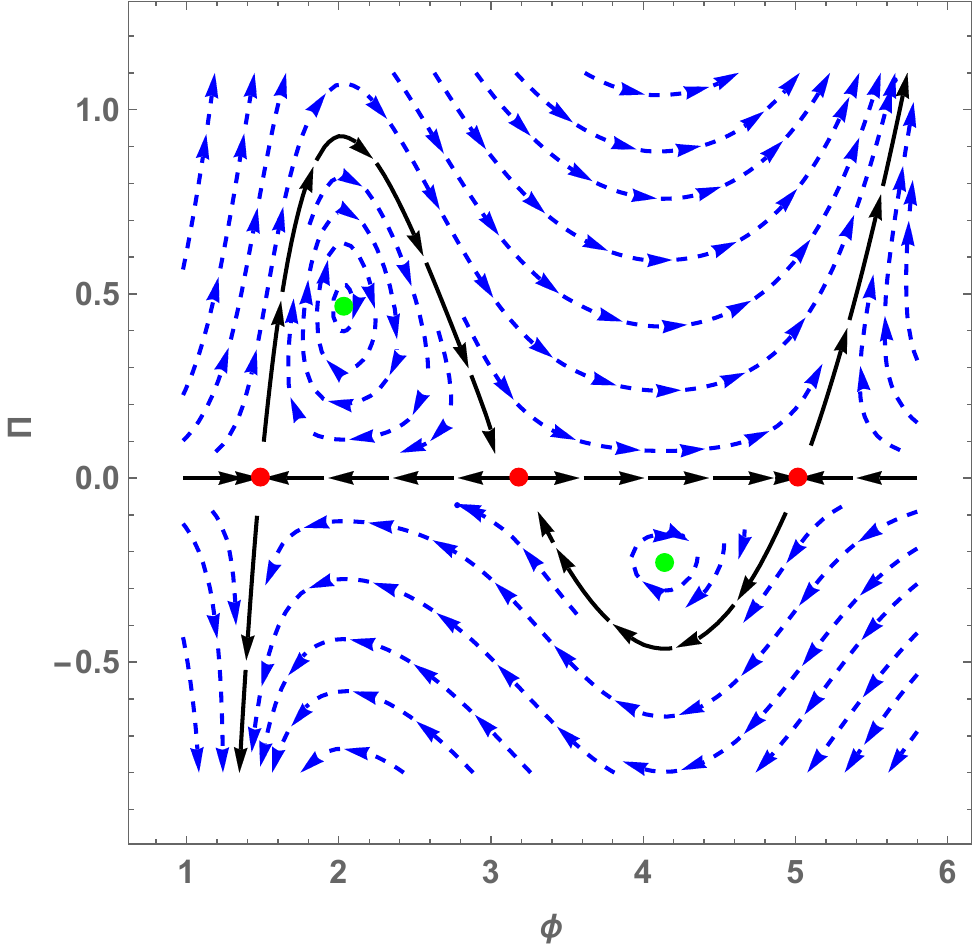}
       \caption{Left panel: Free energy landscape of the RNAdS black holes; Right panel: Hamiltonian flow lines or phase portrait for Hamilton’s equations of motion. The temperature is not equal to the phase transition temperature. The free energy landscape has two wells with different depths.}
    \label{Hamiltonian_flow_line_unequal_depths}
\end{figure}

In Figure \ref{Hamiltonian_flow_line_unequal_depths}, we plot the Hamiltonian flow lines for the case that the free energy landscape has two wells with unequal depths. It is qualitatively similar to that of Figure \ref{Hamiltonian_flow_lines}. From the free energy landscape, one can see that the small black hole is globally stable while the large black hole is locally stable. Under the thermal fluctuations, there is also a state switching process between the two black holes, which is similar to the case at the phase transition point. The corresponding tunneling configuration is analogous to the discussion on the Figure \ref{Hamiltonian_flow_lines}.

\begin{figure}
    \centering
    \includegraphics[width=0.45\textwidth]{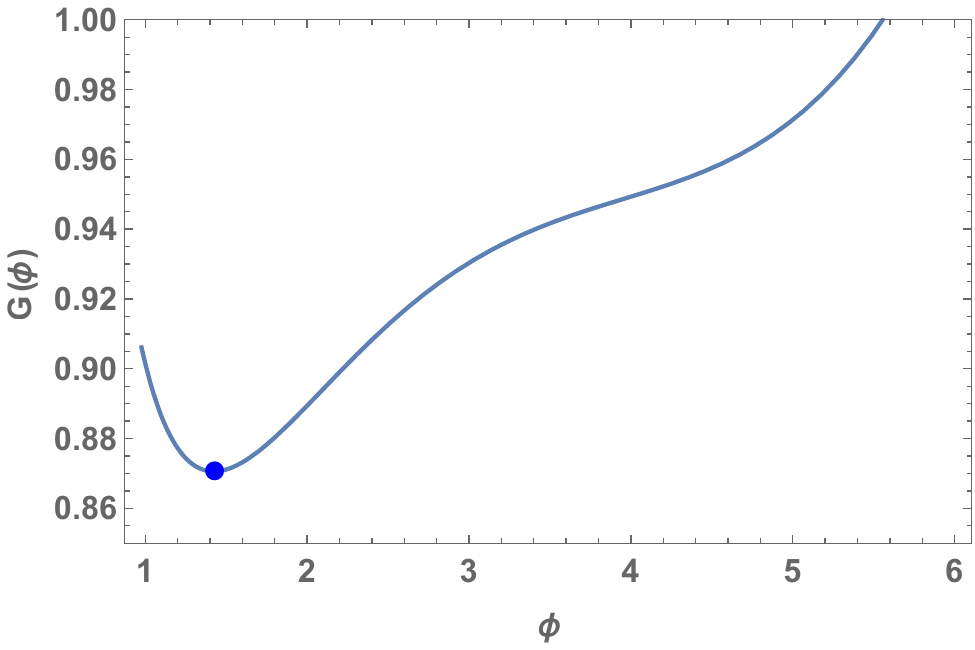}\;\;\;
    \includegraphics[width=0.45\textwidth]{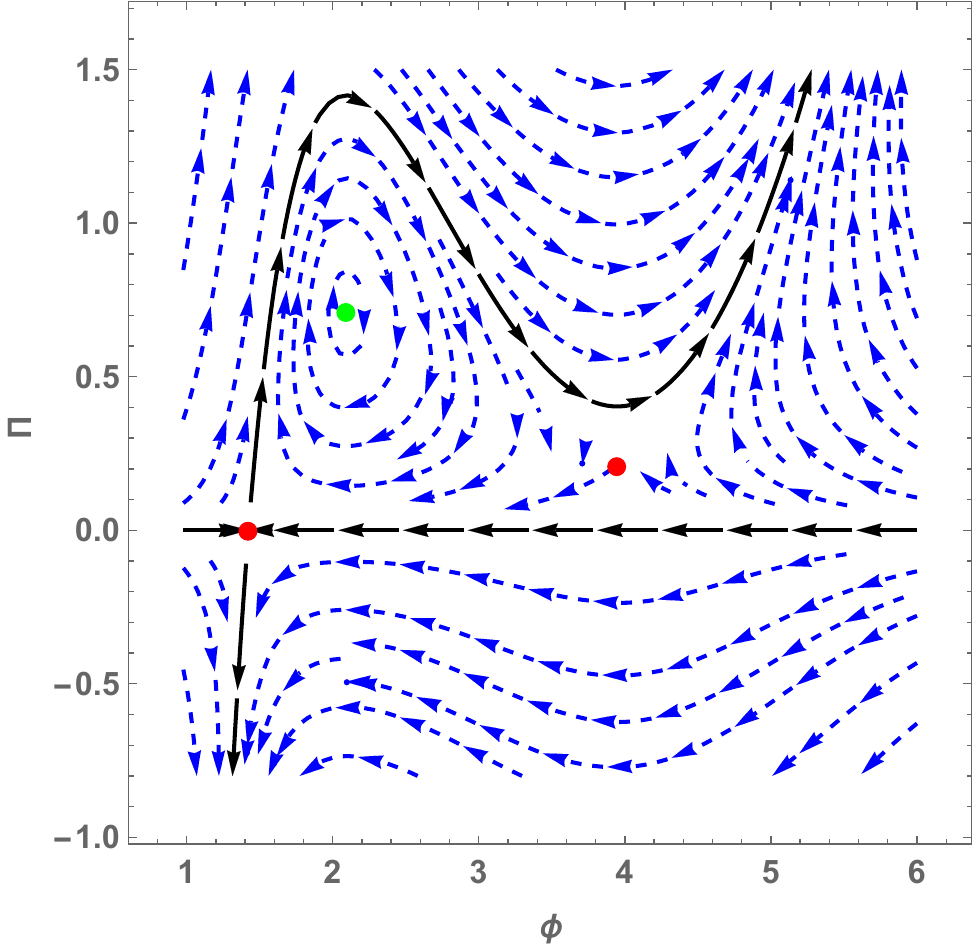}
       \caption{Left panel: Free energy landscape of the RNAdS black holes; Right panel: Hamiltonian flow lines or phase portrait for Hamilton’s equations of motion. The temperature is taken to make sure that the free energy landscape has only a single well.  }
    \label{Hamiltonian_flow_lines_single_well}
\end{figure}

In Figure \ref{Hamiltonian_flow_lines_single_well}, we plot the Hamiltonian flow lines for the case that the landscape has only a single well. In this case, the stationary point on the landscape represents the globally stable black hole state. There are three fixed points in the phase space. The red points are the saddle points and the green point is a center. However, only the left saddle point is relevant to the question under consideration. It represents the stationary black hole on the free energy landscape. This saddle point is unstable in the phase space. In this case, there is no tunneling configuration.

\begin{figure}
    \centering
    \includegraphics[width=0.45\textwidth]{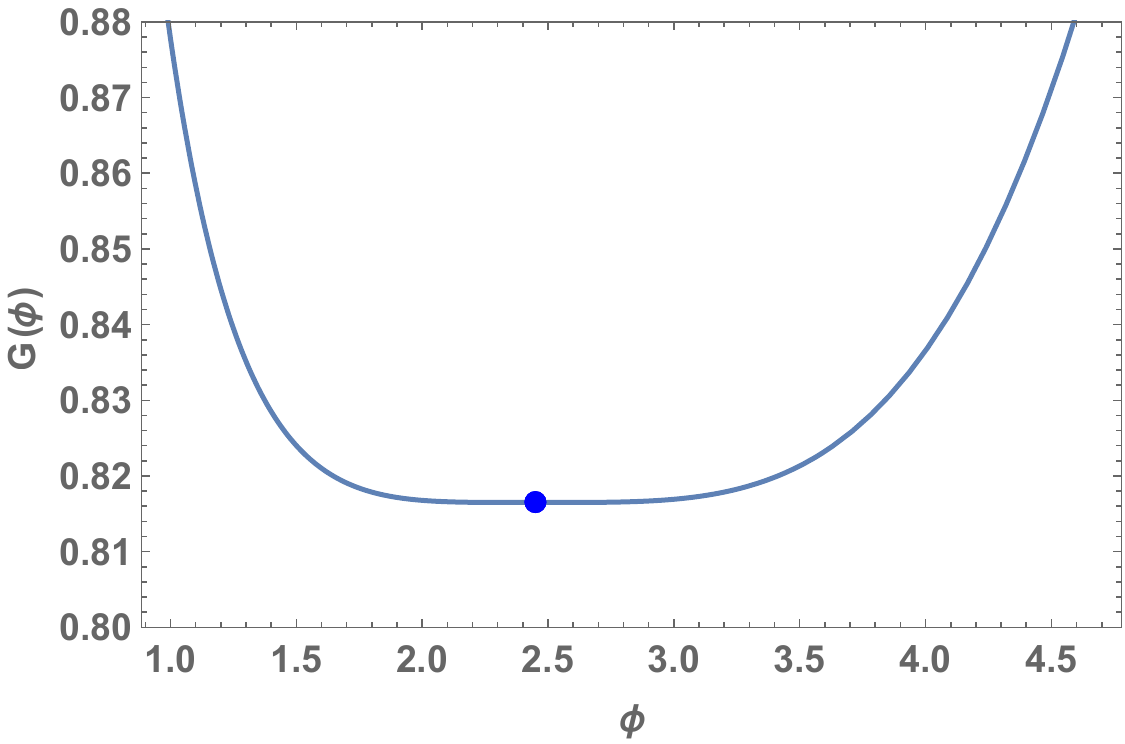}\;\;\;
    \includegraphics[width=0.45\textwidth]{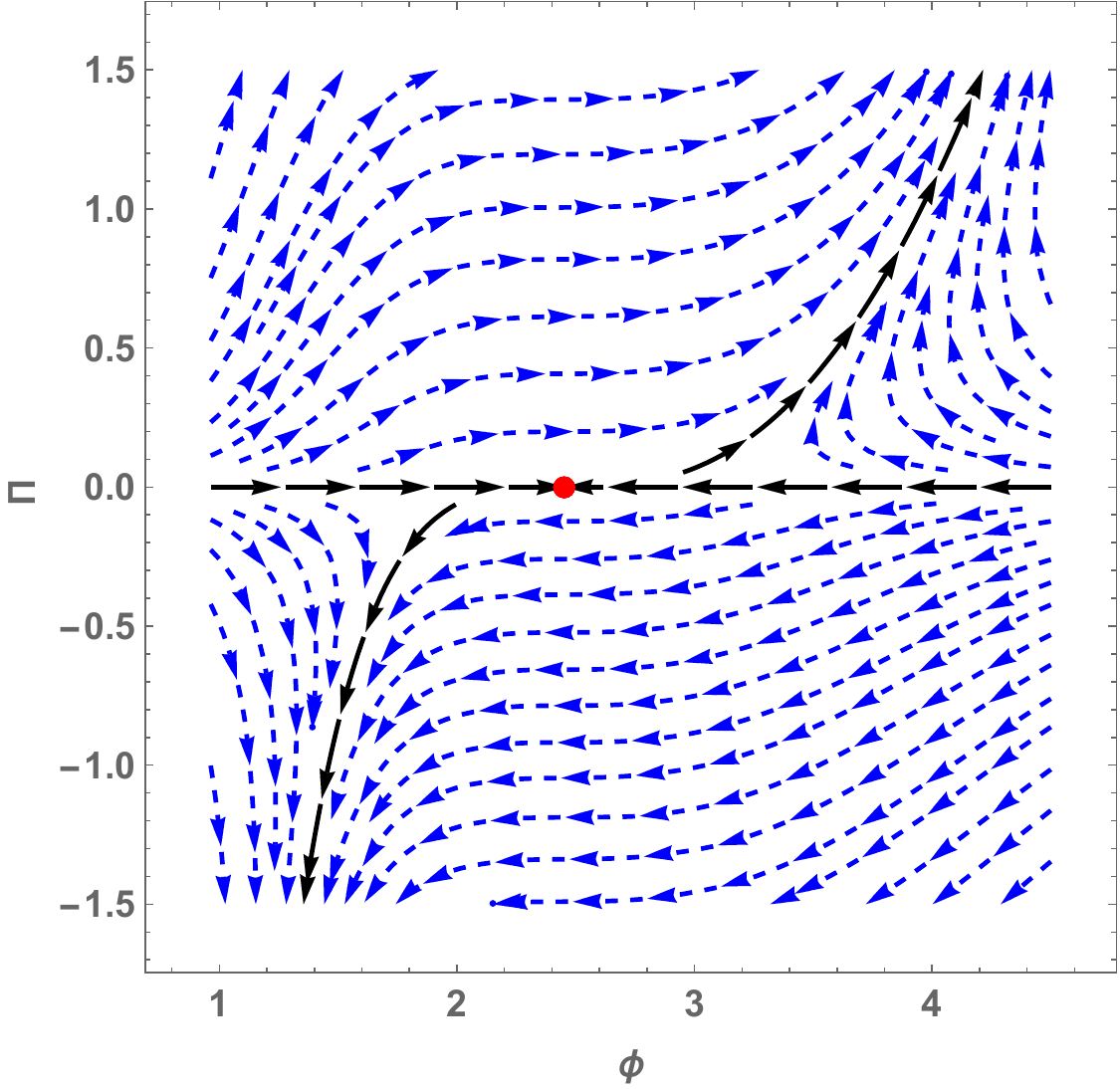}
       \caption{Left panel: Free energy landscape of the RNAdS black hole at the critical point; Right panel: Hamiltonian flow lines or phase portrait for Hamilton’s equations of motion. The temperature and the pressure are taken to be the critical values.  }
    \label{Hamiltonian_flow_critical_point}
\end{figure}

The significance of the RNAdS black hole phase transition in extended phase space is that there exists a critical point analogous to the van der Waals types of liquid-gas phase transition. Here, we also study the Hamiltonian flow lines of the system at the critical point. The critical point for the RNAdS black hole phase transition is determined by the equation 
\begin{eqnarray}
    G'(\phi_c)=G''(\phi_c)=G'''(\phi_c)=0\;,
\end{eqnarray}
from which we can get
\begin{eqnarray}
   && 4 \pi P \phi_c^2-\frac{Q^2}{2 \phi_c^2}-2 \pi T \phi_c+\frac{1}{2}=0\;,\nonumber\\
    && 8\pi P \phi_c+\frac{Q^2}{\phi_c^3} - 2 \pi T=0\;,\nonumber\\
    && 8 \pi  P - \frac{3 Q^2}{\phi_c^4}=0\;,
\end{eqnarray}
where $\phi_c$ is the value of the order parameter for the RNAdS black hole at the critical point. By solving the above equations, one can get the critical point as 
\begin{eqnarray}
    \phi_c=\sqrt{6} Q\;,\quad
    T_c= \frac{1}{3 \sqrt{6} \pi  Q}\;,\quad 
    P_c =\frac{1}{96 \pi  Q^2}\;.
\end{eqnarray}
The free energy landscape of the RNAdS black hole at the critical point is plotted in the left panel of Figure \ref{Hamiltonian_flow_critical_point}. The shape of the free energy landscape is a single well, which is similar to the shape of the free energy landscape we have discussed in Figure \ref{Hamiltonian_flow_lines_single_well}. However, it shows a very flat potential well, which means a dramatically different kinetics near the critical point. The corresponding Hamiltonian flow lines are plotted in the right panel of Figure \ref{Hamiltonian_flow_critical_point}. In this case, there is only one fixed point on the phase space, which represents the only stationary state on the landscape. This is different from the case that was considered in Figure \ref{Hamiltonian_flow_lines_single_well}, where there are three fixed points on the phase space. In addition, there is no tunneling configuration at the critical point also. On $\Pi=0$ flow lines, the fixed point is stable, while on the $\Pi \neq 0$ flow lines, it is unstable.

Note that in the phase space, some of the fixed points of the Hamiltonian equations of motion represent the on-shell black hole solutions. Introduce the vector field $\Phi$ as
\begin{eqnarray}
    \Phi=\left(2D\Pi-\frac{1}{\zeta} G'(\phi), \frac{1}{\zeta} G''(\phi)\Pi \right)\;.
\end{eqnarray}
The fixed points in the phase space are just the singularities of the vector field $\Phi$. Therefore, on-shell black hole solution can be treated as the topological defect of the introduced vector field. However, this is not a universal conclusion because there are fixed points (the green points shown in Figure \ref{Hamiltonian_flow_lines}, Figure \ref{Hamiltonian_flow_line_unequal_depths}, Figure \ref{Hamiltonian_flow_lines_single_well}) that have no physical significance. On the other hand, it is well known that the dynamical stability of the fixed point is relevant to its index or winding number. Here, we find that there is no direct connection between the thermodynamic stability of the black hole solution and the dynamical stability of the fixed point in the phase space. As mentioned, the black hole solutions are all represented as the saddle points in the phase space, which is irrelevant to their thermodynamic stability.

\subsection{Kinetic path in configuration space}

We now consider the tunneling configuration for the small/large RNAdS black hole transition. Based on the dominant paths in the phase space, one can solve the corresponding equation to obtain the dominant kinetic path in the space of the order parameter. According to our previous analysis, it is clear that for the state switching process from the small black hole to the large black hole, the dominant kinetic path is separated into two pieces. The first part is the state switching process from the small black hole to the intermediate black hole, and the second part is the process from the intermediate one to the large black hole.

\begin{figure}
    \centering
    \includegraphics[width=0.45\textwidth]{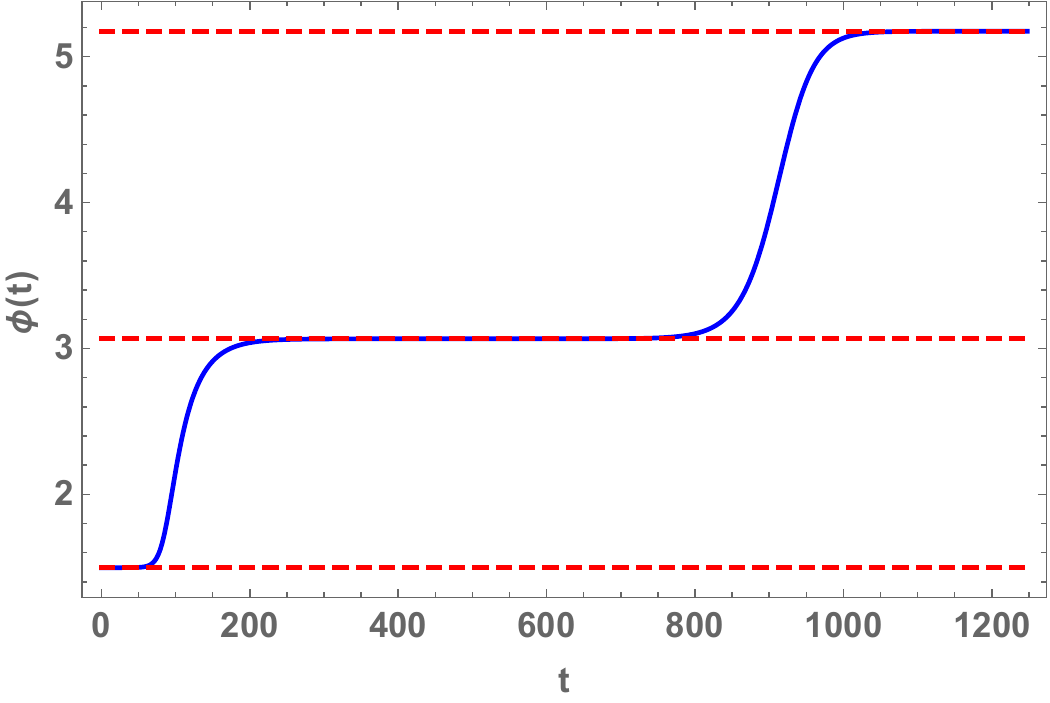}\;\;\;
    \includegraphics[width=0.45\textwidth]{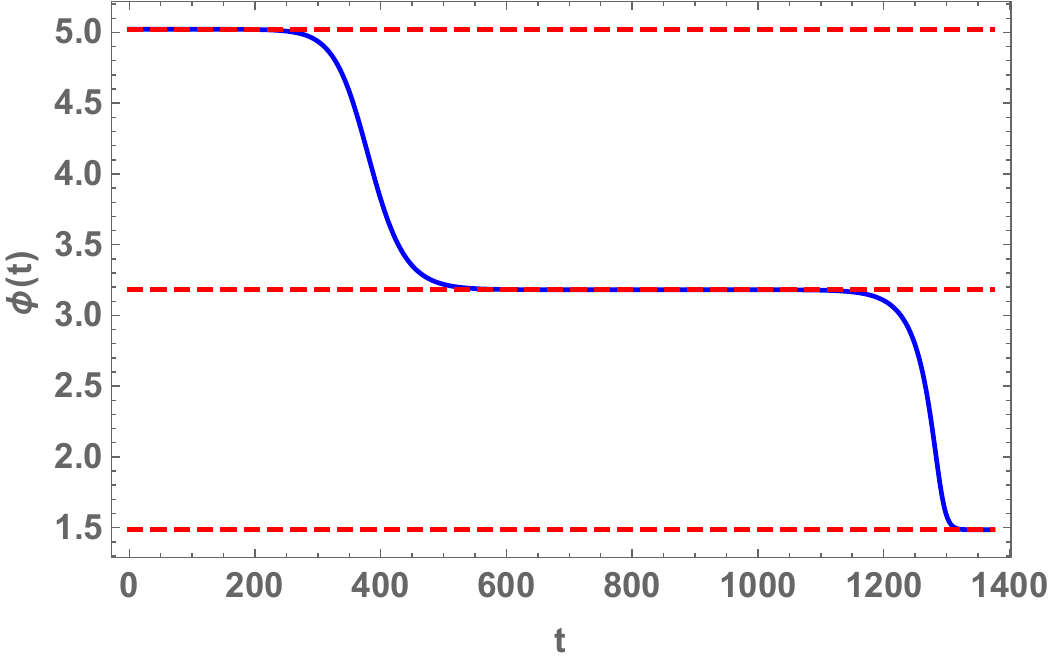}
       \caption{Dominant kinetic paths in the configuration space. Left panel: from the small to the large black hole; Right panel: from the large to the small black hole. The plots are corresponding to Figure \ref{Hamiltonian_flow_lines}, where the two wells on the free energy landscape have the same depth. The three horizontal dashed lines represents $\phi=r_s,r_m$ and $r_l$, respectively. }
    \label{Kinetic_path}
\end{figure}

For the first part, we choose the initial condition as $\phi(t=0)=r_s+\delta$ with $\delta=10^{-5}$ and solve Eq.\eqref{Pineq0}. In this way, we get the kinetic path of the state switching from the small black hole to the intermediate black hole. The criterion that the system reaches the intermediate black hole is given by the difference between the order parameter and the horizon radius of the intermediate state is less than $\delta$, i.e. $\left|\phi(t_1)-r_m\right|<\delta$. When this condition is attained, i.e. the system reaches the top of free energy potential. Then we switch to Eq.\eqref{Pieq0} with the initial condition $\phi(t_1)=r_m+\delta$ to obtain the kinetic path of the state switching from the intermediate black hole to the large black hole. Combining these two paths, we can get the whole path for the kinetic process, which is shown in the left panel of Figure \ref{Kinetic_path}. The kinetic path for the inverse process is also shown in the right panel of Figure.\ref{Kinetic_path}. The initial condition is chosen to guarantee that there is a small derivation from the stationary state. Otherwise, the system will take infinite time to reach another stationary state. For example, we consider the system switching from the $\phi=r_m$ state to the $\phi=r_l$ state along the $\Pi=0$ flow line. The time interval is given by
\begin{eqnarray}
    \Delta t=-\zeta\int_{r_m}^{r_l} \frac{d\phi}{G'(\phi)} \;,
\end{eqnarray}
which is divergent because near the stationary points $\phi=r_m$ and $\phi=r_l$, the asymptotic expansions of the free energy function are quadratic.

\subsection{Kinetic rate}

\begin{figure}
    \centering
    \includegraphics[width=0.4\textwidth]{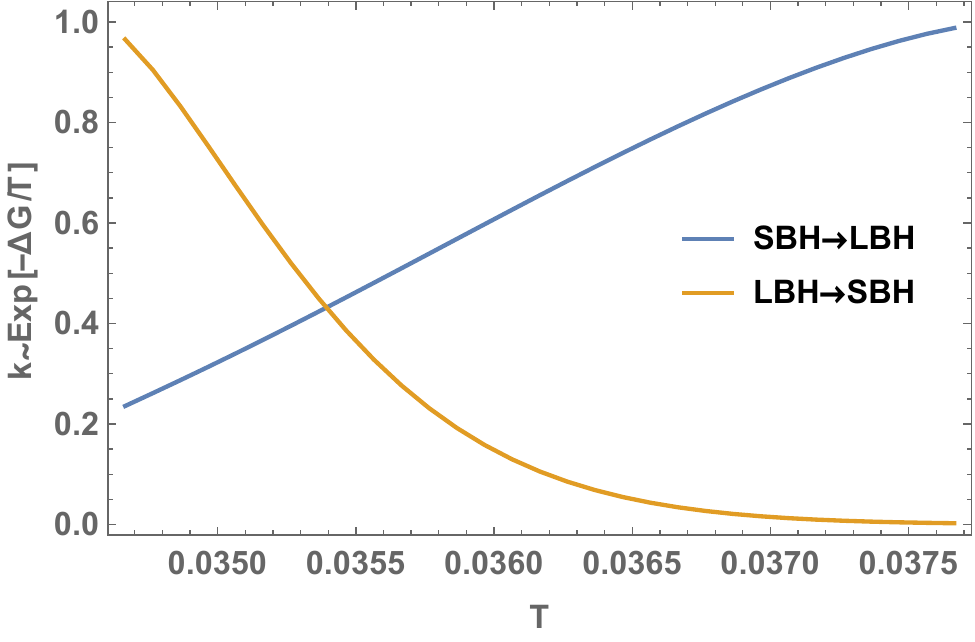}
       \caption{Kinetic rates versus the ensemble temperature. The ensemble temperature is selected to guarantee the free energy landscape has the shape of double well. The two curves represent the rates of the state switching process from the small black hole to the large black hole and its inverse process. The intersection is corresponding to the phase transition point where the two wells on the free energy landscape have the same depth. }
    \label{Kinetic_rate_RNAdS}
\end{figure}

Now we calculate the kinetic rate for the state switching process from the small black hole to the large black hole, i.e. we should evaluate the action in Eq.\eqref{kinetic_rate_PI} 
\begin{eqnarray}
   I= \int_{t_0}^{t} dt' \left(\Pi \dot{\phi}-\mathcal{H}\right)\;.
\end{eqnarray}
Note that the dominant kinetic path of the state switching process corresponds to $\mathcal{H}=0$. In addition, for the kinetic path from the intermediate black hole to the large black hole, $\Pi=0$. Therefore, the nonzero contribution to the action comes from the kinetic path from the small black hole to the intermediate black hole. Then we can get
\begin{eqnarray}
    I=\int_{t_0}^{t_0+t_1}dt \Pi \dot{\phi}
    =\frac{1}{T}\int_{r_s}^{r_m}d\phi G'(\phi)=\frac{\Delta G}{T}\;,
\end{eqnarray}
where $t_0$ is the initial time, $t_0+t_1$ is the time that the order parameter reaches the top of free energy potential and $\Delta G=G(r_m)-G(r_s)$ is the difference of free energy between the intermediate black hole and the small black hole. The final result shows that the action is independent of the time $t_0$ and $t_1$. The kinetic rate is then given by 
\begin{eqnarray}\label{kinetic_rate_eq}
    k\sim \exp\left(-\frac{\Delta G}{T}\right)\;.
\end{eqnarray}
Therefore, the kinetic rate resembles the form of the Kramer rate \cite{Kramers}. The reason is that we have used the assumption that the friction is large enough and the dynamics is in the overdamped regime. This equation means that the kinetic rate is mainly determined by the barrier height on the free energy landscape, which is consistent with our previous results on the kinetics of the black hole phase transition \cite{Li:2021vdp}. For the inverse process, the kinetic rate is also given by the above equation, but the free energy difference is given by $\Delta G=G(r_m)-G(r_l)$. The equation for the kinetic rate also implies that the kinetic process is a fast/slow one for the shallow/deep potential well. We present the numerical results of the kinetic rates in Figure \ref{Kinetic_rate_RNAdS}.

The configurations shown in Figure \ref{Kinetic_path} give the dominant kinetic processes during the black hole phase transition. Because of the existence of the thermal noise, the practical kinetic paths are inherently stochastic. Note that in deriving the Eq.\eqref{kinetic_rate_eq}, we have used the dominant kinetic path and does not take the fluctuation effects around the dominant path into account. When considering these effects, one can get a more precise formula of the kinetic rate \cite{Liu:2021lmr}.

\section{Two further examples}
\label{Sec_V}

In this section, we consider two additional examples, i.e. Hawking-Page phase transition \cite{Li:2020khm} and Gauss-Bonnet black hole phase transition at the triple point \cite{Wei:2021bwy,Li:2023men}. We also point some problems in these two examples.

\subsection{Hawking-Page phase transition}

We now discuss the dominant kinetic path in the phase space for the Hawking-Page phase transition. The basic idea is similar to that used in the last section, but instead utilizing the generalized free energy of the Schwarzschild AdS (SAdS) black hole, which has been calculated in \cite{Li:2020khm} 
\begin{eqnarray}
    G=\frac{\phi}{2}\left(1+\frac{\phi^2}{L^2}\right)-\pi T \phi^2\;.
\end{eqnarray}
Here, the order parameter $\phi$ is the horizon radius of the SAdS black hole, $L$ is the AdS radius and $T$ is the ensemble temperature. The state with $\phi=0$ represents the thermal AdS space. In addition, when $\phi=0$, $G=0$. The generalized free energy is intrinsically the free energy difference between the SAdS black hole and the thermal AdS space.

The phase transition point can be determined by two conditions: (1) Equating the free energies of the AdS space and the SAdS black hole; (2) Ensuring that the derivative of the generalized free energy at the state of the large SAdS black hole is zero. At the phase transition point, the following two equations should be satisfied 
\begin{eqnarray}
    G(\phi_l)&=&\frac{\phi_l}{2}\left(1+\frac{\phi_l^2}{L^2}\right)-\pi T_{HP} \phi_l^2=0\;,\label{G_eq}\\
    G'(\phi_l)&=&\frac{1}{2}+\frac{3\phi_l^2}{2L^2}-2\pi T_{HP} \phi_l=0\;,\label{G_prime_eq}
\end{eqnarray}
where $T_{HP}$ is the Hawking-Page temperature and $\phi_l$ denotes the horizon radius of the large SAdS black hole at the phase transition point. The solution can be easily obtained as
\begin{eqnarray}
    T_{HP}=\frac{1}{\pi L}\;,\;\;\;\phi_l=L\;.
\end{eqnarray}
The radius of the small SAdS black hole can be determined by substituting $T_{HP}=\frac{1}{\pi L}$ into Eq.\eqref{G_prime_eq} and solving the resultant equation, which gives 
\begin{eqnarray}
    \phi_s=\frac{1}{3L}\;.
\end{eqnarray}
It is clear that $G(\phi_s)=\frac{2}{27L}>0$, which implies that the small SAdS black hole is unstable.

\begin{figure}
    \centering
    \includegraphics[width=0.45\textwidth]{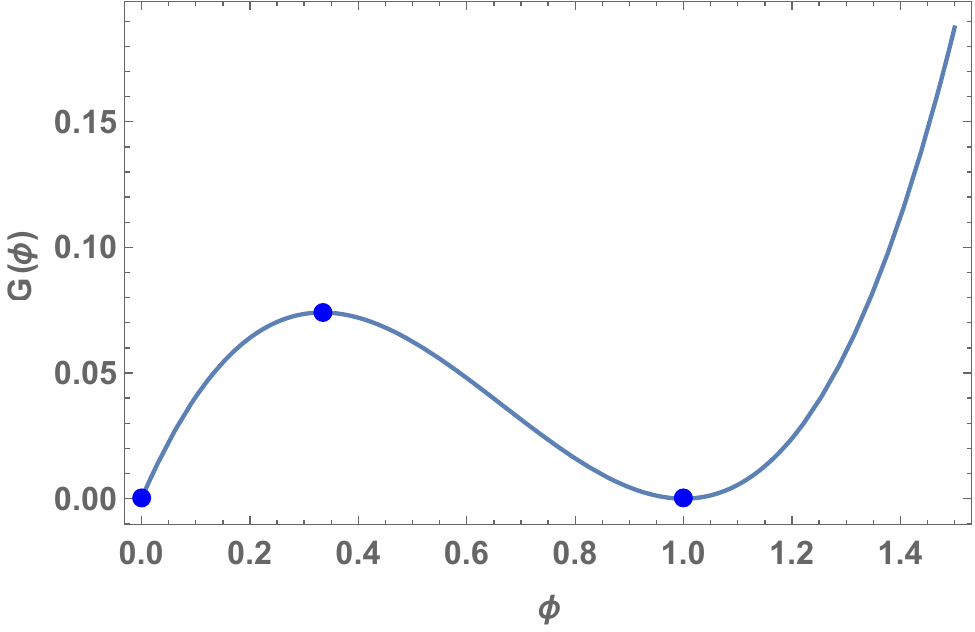}\;\;\;
    \includegraphics[width=0.45\textwidth]{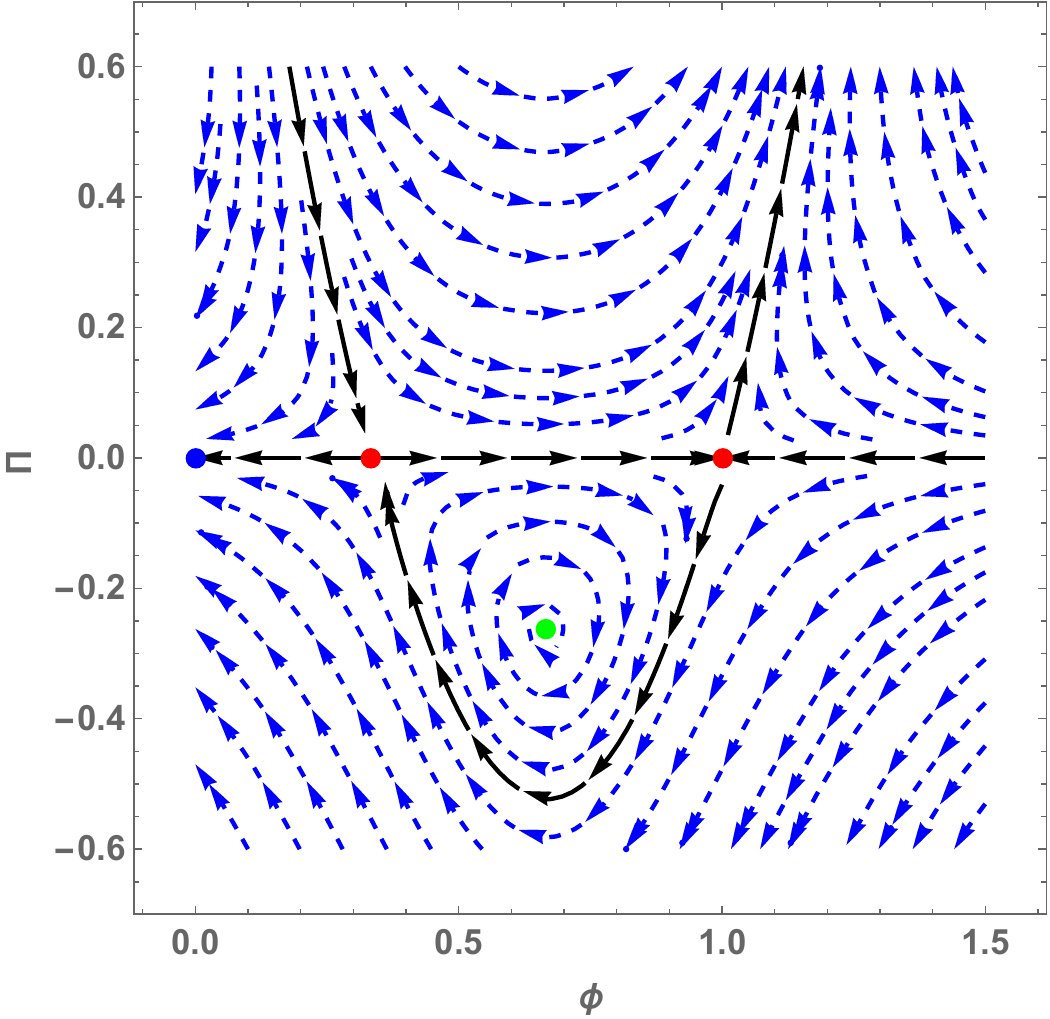}
       \caption{Left panel: Free energy landscape for Hawking-Page phase transition.  Right panel: Hamiltonian flow lines for Hawking-Page phase transition. The AdS radius $L$ is set to be unity. The red points and the green point represent the fixed point while the blue point at $(0,0)$ is not. The two red points correspond to the small and the large SAdS black holes while the blue point is added to denote the thermal AdS space. It seems that there is no flow line corresponding to the state switching process from the thermal AdS space to the large SAdS black hole.}
    \label{Hamiltonian_flow_lines_HP}
\end{figure}

Along the line in the last section, by solving the effective Hamiltonian equations of motion, we can get the Hamiltonian flows as well as the dominant kinetic paths, which are presented in the right panel of Figure \ref{Hamiltonian_flow_lines_HP}. The corresponding free energy landscape is shown in the left panel. In this plot, the ensemble temperature is selected to be the Hawking-Page temperature. In the phase space, there are three fixed points, two points (red) lying in the $\Pi=0$ axial and one point (green) not. As we have discussed, the fixed point with $\Pi\neq 0$ is not relevant to the state switching process associated with Hawking-Page phase transition. The fixed points on the $\phi$ axial corresponds to the small and the large SAdS black holes. The thermal AdS space which is denoted as the blue point in the phase space is not a fixed point of the Hamiltonian flow lines. This is different from the case of RNAdS black hole phase transition, where all the on-shell black hole solutions are the fixed points of the Hamiltonian flow lines. From the example of the RNAdS black hole, we know that the fixed points that are relevant to our discussion satisfy the equation $G'(\phi)=0$. It is obvious that the AdS space with order parameter $\phi=0$ dose not satisfy this equation. Therefore, it cannot be the fixed point of the Hamiltonian flow lines.

The black lines denote the Hamiltonian flow lines with $\mathcal{H}=0$, which are clearly related to the state switching process of Hawking-Page phase transition. The dominant path from the large black hole to the thermal AdS space is starting from the red point on the right (the state of the large black hole), going along with the black line in the lower half plane to the red point on the left (the state of the small black hole), and finally going along with the black line in the $\phi$ axial to reach the blue point (the state of the thermal AdS space). However, the path for the inverse process does not exist. Note that although the free energy of the thermal AdS space is zero, the first derivative of the free energy at the state of the thermal AdS space is not vanishing. This leads to the observation that the thermal AdS space is not a fixed point of the Hamiltonian flow lines. This is the main reason that there is no ``zero energy" flow line that starts from state of the thermal AdS space to the state of the large AdS black hole.

\subsection{Gauss-Bonnet AdS black hole phase transition at the triple point}

We now discuss another example that seems a little exotic. We consider the Gauss-Bonnet AdS black hole in six dimensions, where the free energy landscape has the shape of triple well. The generalized free energy is given by \cite{Wei:2021bwy,Li:2023men}
\begin{eqnarray}
    G=\frac{2\pi }{3}\left(\phi^3+\frac{4}{5}\pi P \phi^5+\alpha \phi +\frac{Q^2}{\phi^3}\right)-\frac{2}{3}\pi^2 T \phi^4 \left(1+\frac{4\alpha}{\phi^2}\right)\;,
\end{eqnarray}
where $P$ is the thermodynamic pressure, $\alpha$ is the Gauss-Bonnet coupling constant, $Q$ is the charge of the black hole and $T$ is the ensemble temperature.

\begin{figure}
    \centering
    \includegraphics[width=0.45\textwidth]{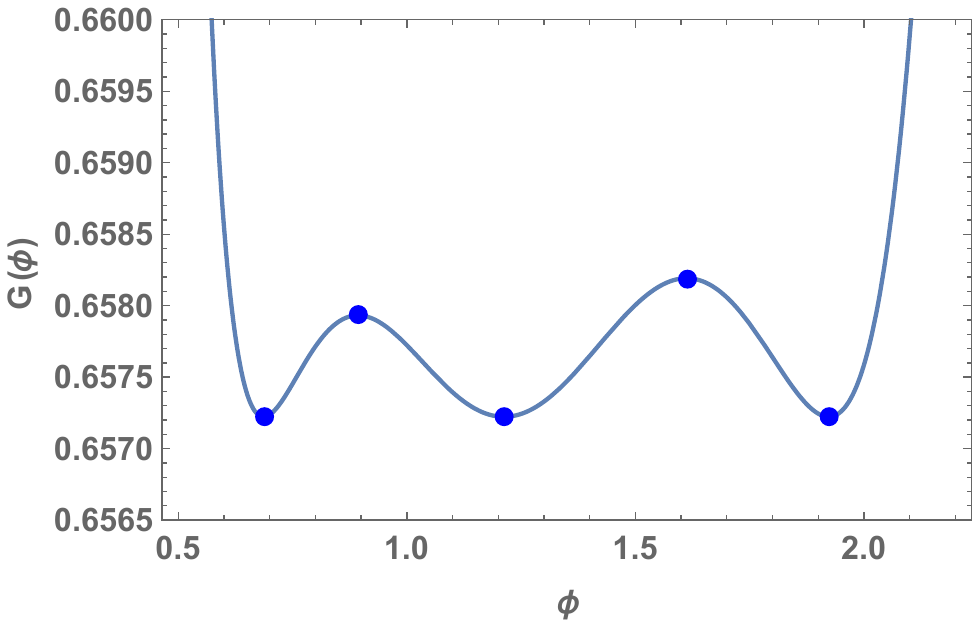}\;\;\;
    \includegraphics[width=0.45\textwidth]{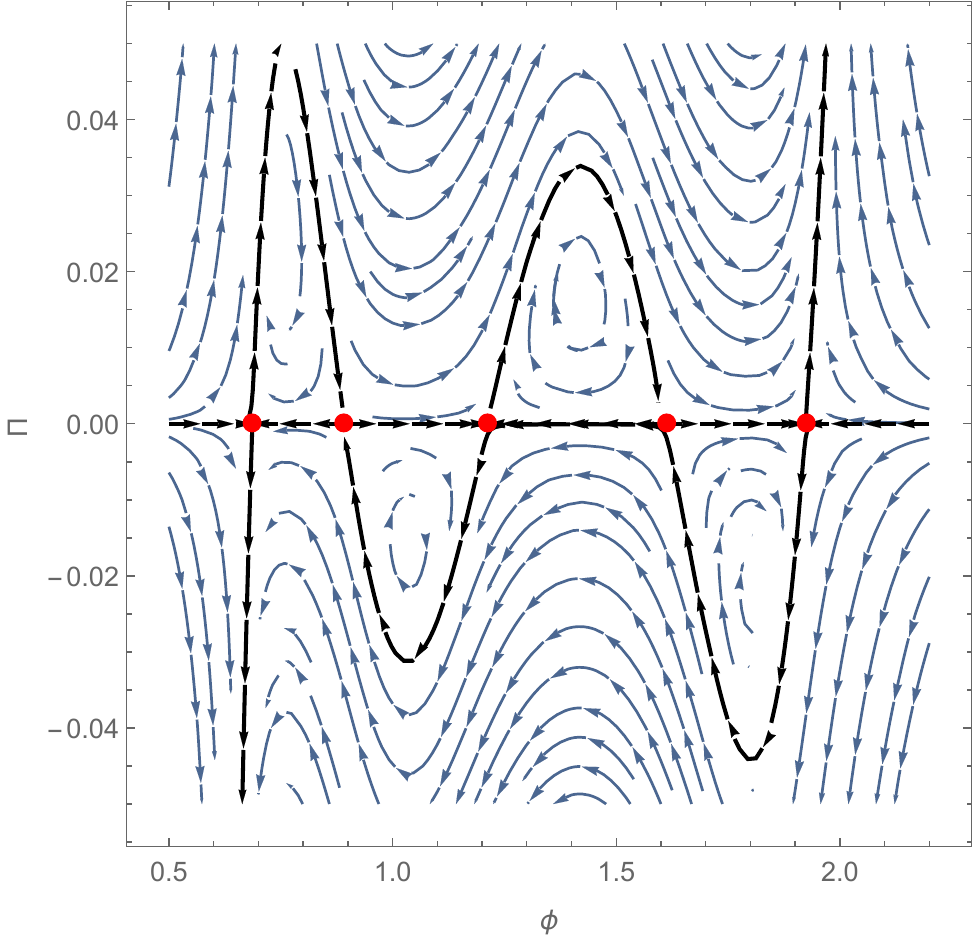}
       \caption{Left panel: Free energy landscape of the Gauss-Bonnet AdS black hole phase transition at the triple point; Right Panel: Hamiltonian flow lines for the phase transition at triple point. In this plot, $\alpha=1.05$, $Q=0.08$, $P=0.0186376$ and $T=0.109523$. The five blue points on the free energy landscape corresponding to the five red points on the phase space are five Gauss-Bonnet AdS black holes with the horizon radii given by $\phi_s=0.687$, $\phi_{us1}=0.893$, $\phi_m=1.213$, $\phi_{us2}=1.612$ and $\phi_l=1.923$ from the smaller value to the larger value.}
    \label{Hamiltonian_flow_lines_GB}
\end{figure}

At the triple point, there are three branches of stable solutions with the order parameters denoted as $\phi_s$, $\phi_m$ and $\phi_l$ and two branches of unstable solutions with the order parameters denoted by $\phi_{us1}$ and $\phi_{us2}$. For the specific charge $Q$ and coupling constant $\alpha$, the pressure and the temperature of the triple point is determined by 
\begin{eqnarray}
    G(\phi_s)=G(\phi_m)=G(\phi_l)\;,\;\;\;G'(\phi_s)=G'(\phi_m)=G'(\phi_l)=0\;.
\end{eqnarray}
By numerically solving these algebraic equations, we can obtain the pressure and the phase transition temperature of the triple point.

At the triple point, the free energy landscape is plotted in the left panel of Figure \ref{Hamiltonian_flow_lines_GB}. The corresponding Hamiltonian flow lines are plotted in the right panel. The black lines, which correspond to $\mathcal{H}=0$, are the flow lines that are relevant to the kinetic process of the state switching. For example, for the state switching process from the small black hole to the large black hole, the dominant kinetic path in the phase space is starting from the point $(\phi_s,0)$, going along the flow line with $\Pi\neq 0$, reaching the point $(\phi_{us1},0)$, then going along the flow line with $\Pi=0$ to reach the point $(\phi_m,0)$, then going along the flow line with $\Pi\neq 0$ to reaching the point $(\phi_{us2},0)$, and finally going along the line with $\Pi=0$ to reach the point $(\phi_l,0)$. Similarly, one can obtain the dominant kinetic path for any pair of locally stable black hole states, which will not be elaborated here.

\begin{figure}
    \centering
    \includegraphics[width=0.4\textwidth]{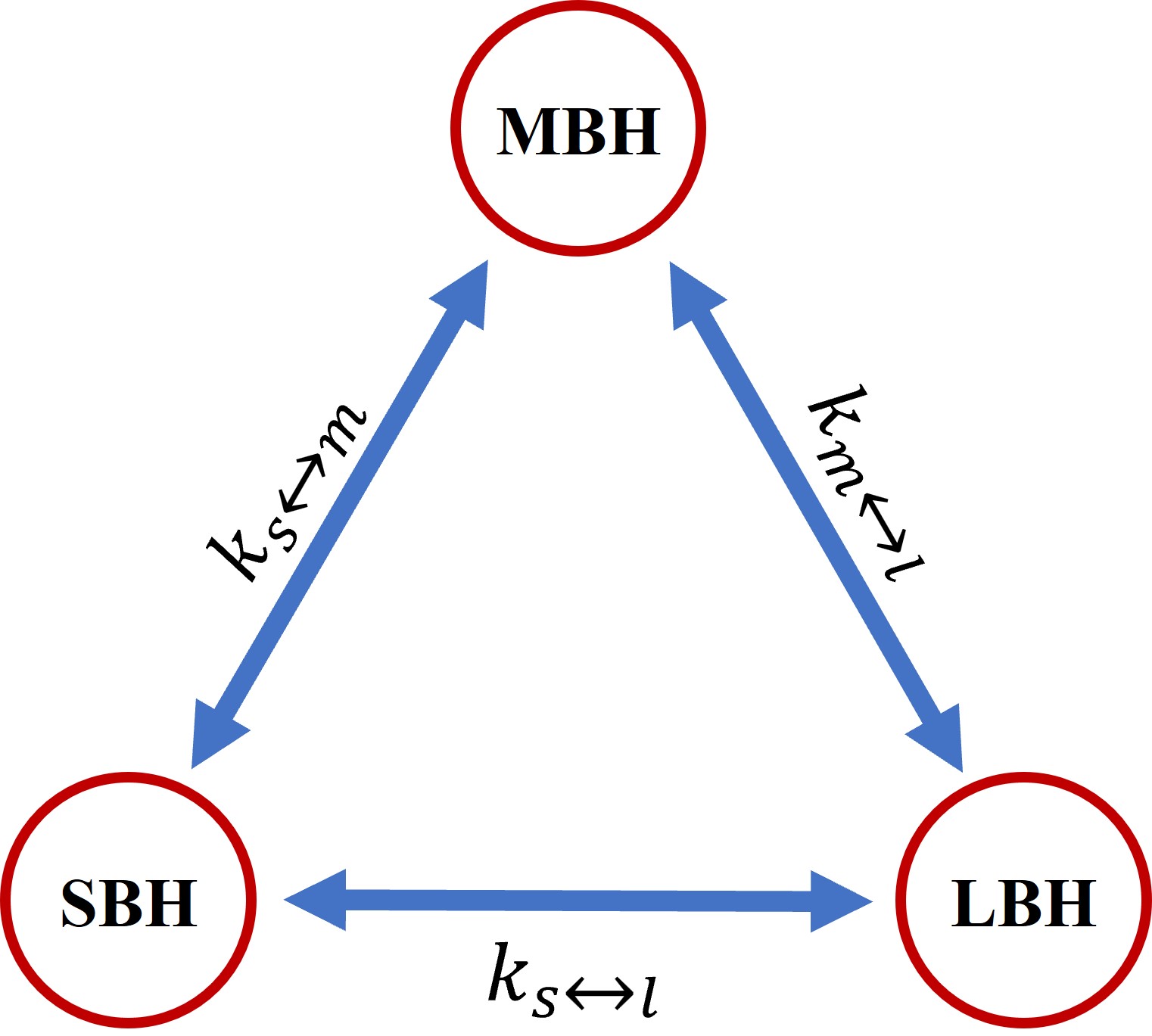}
       \caption{The representation of ``chemical reaction cycle" of the Gauss-Bonnet AdS black hole phase transition at the triple point. SBH, MBH and LBH denote the Gauss-Bonnet AdS black holes with the order parameter $\phi=\phi_s,\phi_m$ and $\phi_l$, respectively.} 
    \label{Kinetic_rate_GB_triple_point}
\end{figure}

For arbitrary dominant kinetic path, only the flow lines with $\Pi\neq 0$ contribute to the kinetic rate. Following the derivation of Eq.\eqref{kinetic_rate_eq}, we can obtain the transition rates in the present case as
\begin{eqnarray}
    k_{s\rightarrow m}&\sim& e^{-\left(G(\phi_{us1})-G(\phi_s)\right)/T}\;,\nonumber\\
    k_{m\rightarrow s}&\sim& e^{-\left(G(\phi_{us1})-G(\phi_m)\right)/T}\;,\nonumber\\
    k_{m\rightarrow l}&\sim& e^{-\left(G(\phi_{us2})-G(\phi_m)\right)/T}\;,\nonumber\\
    k_{l\rightarrow m}&\sim& e^{-\left(G(\phi_{us2})-G(\phi_l)\right)/T}\;,\nonumber\\
    k_{s\rightarrow l}&\sim& e^{-\left(G(\phi_{us1})+G(\phi_{us2})-G(\phi_s)-G(\phi_m)\right)/T}\;,\nonumber\\
    k_{l\rightarrow s}&\sim& e^{-\left(G(\phi_{us2})+G(\phi_{us1})-G(\phi_l)-G(\phi_m)\right)/T}\;.
\end{eqnarray}
For example, $k_{s\rightarrow m}$ represents the kinetic rate of the state switching process from the small black hole with $\phi=\phi_s$ to the intermediate black hole with $\phi=\phi_m$. The last two relations reflect the fact that the switching process from the small/large black hole with $\phi=\phi_s/\phi_l$ to the large/small black hole with $\phi=\phi_l/\phi_s$ must pass through the intermediate black hole state with the order parameter $\phi=\phi_m$. Note that we are consider the phase transition at the triple point, where $G(\phi_s)=G(\phi_m)=G(\phi_l)$. So we have 
\begin{eqnarray}
     k_{s\rightarrow m}= k_{m\rightarrow s}\;,\;\;\;
      k_{m\rightarrow l}= k_{l\rightarrow m}\;,\;\;\;
       k_{l\rightarrow s}= k_{s\rightarrow l}\;.
\end{eqnarray}
In this way, we can formulate a ``chemical reaction cycle" for the state switching between the small, the intermediate and the large Gauss-Bonnet black holes as shown in Figure \ref{Kinetic_rate_GB_triple_point}, which resembles the kinetic cycles of the simplified three species enzyme kinetic model \cite{Qian,Wang}.

Note that in the present framework of free energy landscape, the detailed balance is preserved, which can be seen from the fact that the ratio between the products of the forward rates and the backward rates is equal to one
\begin{eqnarray}
    \frac{k_{s\rightarrow m}k_{m\rightarrow l}k_{l\rightarrow s}}{k_{s\rightarrow l}k_{l\rightarrow m}k_{m\rightarrow s}}=1\;.
\end{eqnarray}
This equation also implies that the time reversal symmetry is preserved for the whole system. It's important to note that the above equation remains valid even when the system deviates from the triple point, as long as there are three wells on the free energy landscape, i.e. the system has three stationary states. In fact, the detailed balance and the time reversal symmetry are closely related concepts in physics, particularly in the context of statistical mechanics. It is generally accepted that detailed balance is a consequence of time reversal symmetry in many physical systems. Therefore, the black hole state switching process in the present framework of free energy landscape is essentially a kind of equilibrium phase transitions.  


\section{Conclusion}
\label{Sec_VI}

In summary, we have studied the kinetics of the black hole phase transition by using the MSRJD functional path integral formalism. The kinetics of phase transition mainly concerns two issues, one is the dominant kinetic path that shows how the state switching process takes place during the phase transition and another is the kinetic rate that reflects how the phase transition occurs. We have shown that the approach of the phase space path integral is a powerful tool to resolve these two issues simultaneously.

Starting from the stochastic Langevin equation that describes the dynamics of state switching process associated with the black hole phase transition, we have derived the MSRJD generating functional and the transition probability in path integral formalism, from which the effective Hamiltonian is extracted. We firstly consider the case that the free energy landscape had the shape of double well. By solving the Hamiltonian equation of motion, we obtain the Hamiltonian flow lines and analyze the dominant kinetic path in the phase space. It is shown that the stationary black hole states on the landscape correspond to the fixed points in the phase space. The kinetic path can connect the different fixed points, which indicates the state switching process in the phase space. Furthermore, the dominant kinetic path in the configuration space is identified for the example of the RNAdS black hole. By using the functional formalism, we also derive the kinetic rates of the state switching processes and reveal the universal conclusion that the kinetic rate is exponentially related to the barrier height between two stationary black hole states on the free energy landscape.

We also consider the case of the single well landscape for the RNAdS black hole. When the system is far from the critical point, there are three fixed points in the phase space, while only one fixed point corresponds to the stationary black hole on the landscape. When the system is at the critical point, the landscape has a very flat potential well and there is only one fixed point in the phase space. In these two cases, there is no state switching, correspondingly no kinetic path present in the phase space.

Finally, we consider two further examples, i.e. Hawking-Page phase transition and Gauss-Bonnet black hole phase transition at the triple point. For the Hawking-Page phase transition, it is shown that while the dominant kinetic path in phase space from the large SAdS black hole to the thermal AdS space is present, there is no kinetic path in the phase space for the inverse process. The reason is that the thermal AdS space is not a fixed point of the Hamiltonian flow lines. For the Gauss-Bonnet AdS black hole phase transition at the triple point, the state switching processes between the small, the intermediate and the large black holes constitute a ``chemical reaction cycle". It is shown that the product of the forward rates is equal to that of the backward rates. This observation suggests that the black hole state switching process within the current framework of the free energy landscape is essentially an kind of equilibrium process.

For future directions, it is interesting to extend the discussion to other types of black hole phase transitions, for example, the black hole phase transitions with multiple critical points \cite{Tavakoli:2022kmo,Yang:2023xzv}.

\appendix
\section{Calculation of Euclidean action}
\label{App_A}

Note that the Euclidean time has the period $\beta$, which is determined by the ensemble temperature via the relation $T=\frac{1}{\beta}$. In other words, we can imagine that there is a thermal bath placed at the spatial infinity. The temperature of the thermal bath is the ensemble temperature $T$. Therefore, the Euclidean version of the metric (\ref{RNAdS_metric}) with the general Euclidean time period describes the Euclidean gravitational instanton with the conical singularity $\Sigma$, which is located at the event horizon. Then, the bulk term has an extra contribution from the conical singularity $\Sigma$ \cite{Fursaev:1995ef}. It can be shown that
\begin{eqnarray}\label{can_R}
\int_{\mathcal{M}} \mathcal{R} =4\pi\left(1-\frac{\beta}{\beta_H}\right)\int_{\mathcal{H}} 1+\int_{\mathcal{M}/\Sigma} \mathcal{R}\;,
\end{eqnarray}
where $\beta_H=1/T_H$ is the inverse Hawking temperature, $\mathcal{H}$ represents the event horizon, and $\mathcal{M}/\Sigma$ represents the regular manifold by excising the conical singularity $\Sigma$. Therefore, we have 
\begin{eqnarray}
    I_{bulk}&=&-\left(1-\frac{\beta}{\beta_H}\right)\frac{\mathcal{A}}{4}-\frac{1}{16\pi} \int_{\mathcal{M}/\Sigma} d^4x \sqrt{g} \left( R +\frac{6}{L^2}-F_{\mu\nu}F^{\mu\nu}\right)\nonumber\\&=&
    -\left(1-\frac{\beta}{\beta_H}\right)\frac{\mathcal{A}}{4}+\frac{1}{16\pi} \int_{\mathcal{M}/\Sigma} d^4x \sqrt{g} \left(\frac{6}{L^2}+F_{\mu\nu}F^{\mu\nu}\right)\;,
\end{eqnarray}
where $\mathcal{A}=4\pi r_+^2$ is the horizon area and we have used the fact that $R=-\frac{12}{L^2}$ for the RNAdS metric. It can be calculated that
\begin{eqnarray}
    \frac{1}{16\pi} \int_{\mathcal{M}/\Sigma} d^4x \sqrt{g} \left(\frac{6}{L^2}+F_{\mu\nu}F^{\mu\nu}\right)&=& \frac{\beta}{4} \int_{r_+}^{r_c}r^2  dr \left(\frac{6}{L^2}+\frac{2Q^2}{r^4}\right)
    \nonumber\\&=&
    \frac{\beta}{2}\left[ 
    \frac{1}{L^2}\left(r_c^3-r_+^3\right)-Q^2\left(\frac{1}{r_c}-\frac{1}{r_+}\right)\right]\;,
\end{eqnarray}
where $r_c$ is the radius of the boundary $\partial\mathcal{M}$. Here we choose the boundary $\partial\mathcal{M}=S_1\times S_2$, $S_2$ being a 2-sphere with a large radius, which will be sent to infinity in the final step.

The outpointing unit normal vector $n^{\mu}$ for a hypersurface with fixed $r$ is given by $\left(0,\sqrt{f(r)},0,0\right)$. The extrinsic curvature $K$ of the hypersurface can be calculated as 
\begin{eqnarray}
    K=\nabla_{\mu}n^{\mu}=\frac{1}{r^2}\partial_r\left(r^2\sqrt{f}\right)=\frac{f'}{2\sqrt{f}}+\frac{2}{r}\sqrt{f}\;.
\end{eqnarray}
The Gibbons-Hawking surface term can be calculated as 
\begin{eqnarray}
    -\frac{1}{8\pi} \int_{\partial\mathcal{M}} d^3 x\sqrt{h} K&=&-\frac{\beta}{2}\left.\left(\frac{1}{2}r^2f'(r)+2rf(r)\right)\right|_{r=r_c}\nonumber\\
   &=& -\frac{\beta}{2}\left(\frac{3r_c^3}{L^2}+2r_c-3M+\frac{Q^2}{r_c}\right)\;.
\end{eqnarray}
The electromagnetic surface term is given by 
\begin{eqnarray}
    -\frac{1}{4\pi} \int_{\partial\mathcal{M}} d^3 x\sqrt{h} n_{\mu} F^{\mu\nu} A_{\nu}=\frac{\beta Q^2}{r_c}\;.
\end{eqnarray}
By noting that $\mathcal{R}=\frac{2}{r_c^2}$ and $\mathcal{R}_{ab}\mathcal{R}^{ab}=\frac{2}{r_c^4}$, the counterterm action is given by 
\begin{eqnarray}
   I_{ct}&=&\frac{1}{8\pi} \int_{\partial\mathcal{M}} d^3 x
    \sqrt{h}\left[\frac{2}{L}+\frac{L}{2}\mathcal{R} -\frac{L^3}{2} \left( \mathcal{R}_{ab}\mathcal{R}^{ab}-\frac{3}{8}\mathcal{R}^2\right) \right] \nonumber\\
    &=&\frac{\beta}{2}r_c^2\sqrt{f(r_c)}\left[\frac{2}{L}+\frac{L}{r_c^2}-\frac{L^3}{4r_c^4} \right]\nonumber\\
    &=&\frac{\beta}{2}\left[\frac{2r_c^3}{L^2}+2r_c-2M+\frac{Q^2}{r_c}+\mathcal{O}\left(\frac{1}{r_c^3}\right) \right]\;,
\end{eqnarray}
where we have used the series expansion of $r_c$ at the infinity. 

Taking all the contribution into account and sending $r_c$ to infinity, we can finally get the Euclidean action as 
\begin{eqnarray}
    I_E&=&-\left(1-\frac{\beta}{\beta_H}\right)\frac{\mathcal{A}}{4}+\frac{1}{2}\beta M -\frac{\beta r_+^3}{2L^2}+\frac{\beta Q^2}{2r_+}\nonumber\\
    &=&\beta M-S\;.
\end{eqnarray}
In the last step we have used the Smarr relation
\begin{eqnarray}
    M=2T_H S-2VP+\Phi Q\;,
\end{eqnarray}
where the thermodynamic volume $V$ and the electromagnetic potential $\Phi$ measured at infinity with respect to the horizon are given by 
\begin{eqnarray}
    V=\frac{4}{3}\pi r_+^3\;,\;\;\Phi=\frac{Q}{r_+}\;.
\end{eqnarray}
Note that in Eq.\eqref{action}, we have replaced the mass $M$ and the entropy $S$ by using the expressions \eqref{mass} and \eqref{entropy}. In this way, the Euclidean action $I_E=\beta M-S$ can be completely expressed as the function of the order parameter $r_+$.

\section{Time-dependent Ginzburg-Landau model}
\label{App_B}

Mean-field theory is shown to be a powerful tool to study the collective behaviors of many physical systems. In this appendix, we briefly review a simple model, which is often referred to as the time-dependent Ginzburg-Landau model \cite{Hohenberg}, that is usually utilized to describe the dissipative dynamics of a nonconserved order parameter. When the timescales of the fast and the slow dynamical variables can be separated distinctly, the non-Markovian effects from the thermal environment can be neglected appropriately. In this way, one can set up a stochastic dynamical equation to investigate the dissipative dynamics of the system. However, the rigorous derivation of the Langevin equation should be given by using the projection operator method.

The free energy functional of the system can be often written as \cite{Hohenberg} 
\begin{eqnarray}
    G[\psi]=\int d\vec{r} \left[g(\psi)+\frac{1}{2} K \left(\nabla\psi\right)^2 \right]\;,
\end{eqnarray}
where $\psi=\psi(\vec{r},t)$ is the order parameter of the system depending on the space and time. The local free energy density is denoted as $g(\psi)$. The parameter $K$ can be viewed as the effective ``mass".   

The equation that models the dissipative (over-damped) relaxation of the system to its free-energy minimum can be given by \cite{Hohenberg}
\begin{eqnarray}\label{model_A}
    \frac{\partial \psi(\vec{r},t)}{\partial t}= -\Gamma \frac{\delta G[\psi]}{\delta \psi} +\theta(\vec{r},t)\;,
\end{eqnarray}
where $\frac{\delta G[\psi]}{\delta \psi}$ represents the functional derivative of the free energy $G$ with respect to the order parameter $\psi$, and $\Gamma$ denotes the inverse damping coefficient. The noise term $\theta(\vec{r},t)$ in the above equation is usually spacetime dependent and satisfies the conventional fluctuation-dissipation relation
\begin{eqnarray}
   && \langle \theta(\vec{r},t) \rangle=0\;,\\
   && \langle \theta(\vec{r},t)\theta(\vec{r}',t') \rangle=2\Gamma T \delta(\vec{r}-\vec{r}')\delta(t'-t)\;,
\end{eqnarray}
where we have set the Boltzmann constant $k_B$ to unity for later convenience.

This model is also referred to as Model A of order-parameter kinetics, as discussed by Hohenberg and Halperin in the theory of dynamical critical phenomena \cite{Hohenberg}. At late time, the system will reach the Boltzmann equilibrium distribution at the temperature $T$. This can be justified by writing down the time evolution equation of the probability distribution $P[\psi, t]$, which is also known as the functional Fokker-Planck equation 
\begin{eqnarray}
    \frac{\partial P}{\partial t}= \Gamma \int d\vec{r} \frac{\delta}{\delta \psi}\left[ \frac{\delta P}{\delta\psi} + P \frac{\delta G}{\delta \psi} \right]\;.
\end{eqnarray}
For the time independent solution, it gives the equilibrium Boltzmann distribution determined solely by the free energy functional 
\begin{eqnarray}
    P_{eq}(\psi)=\frac{1}{Z} \exp\left(-G[\psi]\right)\;,
\end{eqnarray}
where $Z=\int \mathcal{D}\psi \exp\left(-G[\psi]\right) $ is the normalization constant.

This model was widely used to describe the collective behavior of the system near the critical point. The kinetic model of black hole state switching was much inspired by the time dependent Ginzburg-Landau model, but with a rather simple setup that the order parameter $\phi$ is only the function of the time $t$, which reduces the spatially inhomogenous equation \eqref{model_A} to an ordinary differential equation.

\end{document}